\documentclass[12pt, centerh1]{article}
\usepackage{natbib}

\usepackage{booktabs}

\usepackage{lettrine}

\usepackage{enumitem}
\setlist[itemize]{noitemsep}

\usepackage{hyperref}

\usepackage{placeins,textgreek,colonequals}
\usepackage{amsmath,tabstackengine,longtable}
\usepackage{amsfonts,rotating}
\usepackage{graphicx,colonequals}
\usepackage{caption,multirow}
\usepackage{subcaption}
\usepackage{pbox}
\usepackage[utf8]{inputenc}
\usepackage[english]{babel}
\usepackage{amsthm}
\usepackage{subcaption,lscape}

\newcommand{\vecx}{\mathbf{x}}
\newcommand{\vecy}{\mathbf{y}}
\newcommand{\vecY}{\mathbf{Y}}

\newcommand{\vecX}{\mathbf{X}}

\newcommand{\vecw}{\mathbf{w}}
\newcommand{\vecz}{\mathbf{z}}

\newcommand{\bs}{\boldsymbol}

\newcommand{\mvn}{\text{MVN}}

\title{Clustering Airbnb Reviews}
\author{Yang Tang and Paul D. McNicholas}
\date{\small Department of Mathematics \& Statistics, McMaster University, Ontario, Canada.}

\begin{document}
\maketitle
\begin{abstract} 
In the last decade, online customer reviews increasingly exert influence on consumers' decision when booking accommodation online. The renewal importance to the concept of word-of mouth is reflected in the growing interests in investigating consumers' experience by analyzing their online reviews through the process of text mining and sentiment analysis. A clustering approach is developed for Boston Airbnb reviews submitted in the English language and collected from 2009 to 2016. This approach is based on a mixture of latent variable models, which provides an appealing framework for handling clustered binary data. We address here the problem of discovering meaningful segments of consumers that are coherent from both the underlying topics and the sentiment behind the reviews. A penalized mixture of latent traits approach is developed to reduce the number of parameters and identify variables that are not informative for clustering. The introduction of component-specific rate parameters avoids the over-penalization that can occur when inferring a shared rate parameter on clustered data. We divided the guests into four groups -- property driven guests, host driven guests, guests with recent overall negative stay and guests with some negative experiences. 

\end{abstract}

\section{Introduction} \label{Intro}
The advent of the sharing economy has changed consumer behaviours dramatically in recent years. Airbnb is the world's largest home sharing platform with more than $800,000$ listings in more than $34,000$ cities. Previous studies have primarily relied on traditional quantitative methods using the Airbnb ratings and/or conduct surveys to investigate guests' experience. For example, \citet{guttentag18} design a study to understand the motivation why so many tourists choose this novel service instead of traditional accommodation options. The study involved an online survey completed by more than 800 consumers who had stayed in Airbnb accommodation during 2014--2015. They divided the respondents into five segments: Money Savers, Home Seekers, Collaborative Consumers, Pragmatic Novelty Seekers, and Interactive Novelty Seekers. However, consumers' reviews already contain rich information and play perhaps the most crucial role in capturing consumer satisfaction as well as opinions about their accommodation and hosts for Airbnb websites \citep{he13, small14, tussyadiah17}. \citet{cheng19} investigate the attributes that influence Airbnb users' experiences by analysing the reviews through the process of text mining, e.g., topic models and individual concept's likelihood scores, and sentiment analysis. They find three key attributes of an Airbnb experience: location, amenities and host. However, due to the nature of these models, this study viewed Airbnb users as homogenous, rather than as members of potential market segments which can offer valuable marketing insights for Airbnb, its hosts, and competing accommodation firms.

Latent Dirichlet allocation \cite[LDA;][]{blei03} is one of the most popular methods among probabilistic approaches for bag of words analysis that automatically discovers topics in text documents. Each document can simultaneously belong to several $G$ uncorrelated topics (clusters). Later on, the correlated topic model \cite[CTM;][]{lafferty06} and the relational topic model \cite[RTM;][]{chang09} were developed to overcome the limitation of taking into account possible topic correlations. While topic models can be instrumental in text mining, it is unnecessary and hard to interpret when clustering Airbnb reviews, especially when the goal is to divide consumers into homogeneous groups that are internally similar in a meaningful way. To achieve the goal of consumer segmentation using bag of words model, and because of the lack of a suitable method, we propose here a new approach based on a mixture of latent trait models. The continuous latent variables allow us to find underlying ``topics'' and the group structure can find homogenous clusters. 

Recent work on the analysis of clustered binary data via mixtures of latent trait models includes the approaches of \cite{muthen06}, \cite{vermunt07}, \cite{browne12}, and \cite{gollini14}. A problem that arises with high-dimensional binary data is the large number of model parameters; consequently, interest in penalized latent variable models for binary data has recently been increasing \citep[see][for examples]{houseman07, desantis08}. \citet{houseman07} propose a penalized item response theory model with univariate traits and penalize the item-response slopes with ridge penalties. However, their approach does not take into account the potential group structure of the data and Gauss-Hermite quadrature is required to approximate the likelihood. \citet{desantis08} develop a penalized latent class model to facilitate analysis of high-dimensional ordinal data. A ridge penalty is introduced to the feature-based parameterization of class-specific response probabilities to stabilize the maximum likelihood estimation. Because the ridge penalty does not encourage sparsity, we search for penalty function with sharp densities spike at zero. Both methods require a model selection criterion, such as the Bayesian information criterion \citep[BIC;][]{schwarz78} to choose the rate parameter. Therefore, they adopt a shared rate parameter, which can cause over-penalization on clustered data, to avoid an exhaustive search. 

For these reasons, a penalized mixture of latent trait models (PMLTM) is proposed for clustered binary data such as Airbnb data. The data are assumed to have been generated by a mixture of latent trait models \citep{gollini14} and we shrink the slope parameters using a gamma-Laplace penalty function \citep{taddy13}. The PMLTM model enables us to encourage sparsity in estimating the slope parameters. The result is a considerable reduction in the number of free parameters as well as automatic variable selection. Moreover, the component-specific independent rate parameter avoids the over-penalization that can occur when inferring a shared rate parameter on clustered data. The newly developed variational expectation-maximization (VEM) algorithm \citep{tipping992, gollini14} provides closed form estimates for model parameters and avoids intensive searches of the rate parameters through a model selection criterion, e.g., the BIC.

\section{Penalized Mixture of Latent Trait Models}\label{sec:method}
\subsection{Overview} \label{sec:overview}
Assume that each observation $ \vecx_i$, $i=1,\ldots, n$ comes from one of the $G$ components and use $\vecz_i=(z_{i1},\ldots, z_{iG})'$ to identify the component membership, where $z_{ig}=1$ if observation $i$ is in component $g$ and $z_{ig}=0$ otherwise. The conditional distribution of $\vecX_i$ in component $g$ is a latent trait model and takes the form
\begin {equation} \label{eq:2.1}
p(\vecx_i|\bs \Theta)=\sum_{g=1}^G\eta_gp(\vecx_i|\bs \theta_g)=\sum_{g=1}^G\eta_g\int \limits_{\vecY_i} p(\vecx_i|\vecy_{i},\bs \theta_g)p(\vecy_{i})d\vecy_{i},
\end{equation}
where
\begin {equation*}
 p(\vecx_i|\vecy_{i}, \bs \theta_g)=\prod_{m=1}^M\{\pi_{mg}(\vecy_{i})\}^{x_{im}}\{1-\pi_{mg}(\vecy_{i})\}^{1-x_{im}}
\end{equation*}
and the response function for each categorical variable in each component is
\begin {equation} \label{eq:2.2}
\pi_{mg}(\vecy_{i})=p(x_{im}=1|\vecy_{i}, \bs \theta_g)=\frac{1}{1+\exp\{-(\alpha_{mg}+\vecw_{mg}'\vecy_{i})\}},
\end{equation}
where $\alpha_{mg}$ and $\vecw_{mg}$ are the model parameters and the multivariate latent variable $\vecY_{i}\sim \text{MVN}(\bs 0, \mathbf{I}_D)$.
Under this model, %these conditions, the approach represents a generalization of the {\color{red}latent class model} where 
observations are not necessarily conditionally independent given the group memberships. In fact, the observations within groups are modelled using a latent trait analysis model and thus dependence is accommodated. This model is known as the MLTA\citep[see][]{gollini14}.

\subsection {Penalized MLTA Models via Non-Convex Penalties}\label{sec:nonconvexpenalties}
A potential drawback of the MLTA for high-dimensional data is its large number of parameters. In particular, the model in \eqref{eq:2.1} involves $(G-1)+GM+G[MD- D(D-1)/2]$ free parameters, of which $G[MD- D(D-1)/2]$ are from $\vecw_{mg}$, for $m=1, \ldots,M$ and $g=1, \ldots G$. 
To reduce the number of free parameters, we propose a penalized log-likelihood of the form
\begin{equation}\label{eq:2.3}
Q(\bs \Theta)=l(\bs \Theta)-C(\bs \Theta),
\end{equation}
where $l(\bs \Theta)$ is the log-likelihood of \eqref{eq:2.1} and $C(\bs \Theta)$ is a penalty term. Similar to the least absolute shrinkage and selection operator (LASSO) penalty for regression \citep{tibshirani96}, we propose use of a heavy-tailed and sparsity-inducing independent Laplace prior for each coefficient $\vecw_{mg}$. To account for uncertainty about the appropriate level of component-and-variable-specific regularization, each Laplace rate parameter $\lambda_{mg}$ is left unknown with a gamma hyperprior. Thus, 
\begin{equation}\label{eq:2.4}
\pi(\vecw_{mg}, \lambda_{mg})=\frac{r^s}{\Gamma(s)}\lambda_{mg}^{s-1} \exp (-r\lambda_{mg})\prod_{d=1}^D\frac{\lambda_{mg}}{2}\exp(-\lambda_{mg}|w_{dmg}|),
\end{equation}
for $s,r>0$.

\begin{figure}[htb]
        \centering
    \vspace{-0.15in}
      \includegraphics[width=0.8\textwidth]{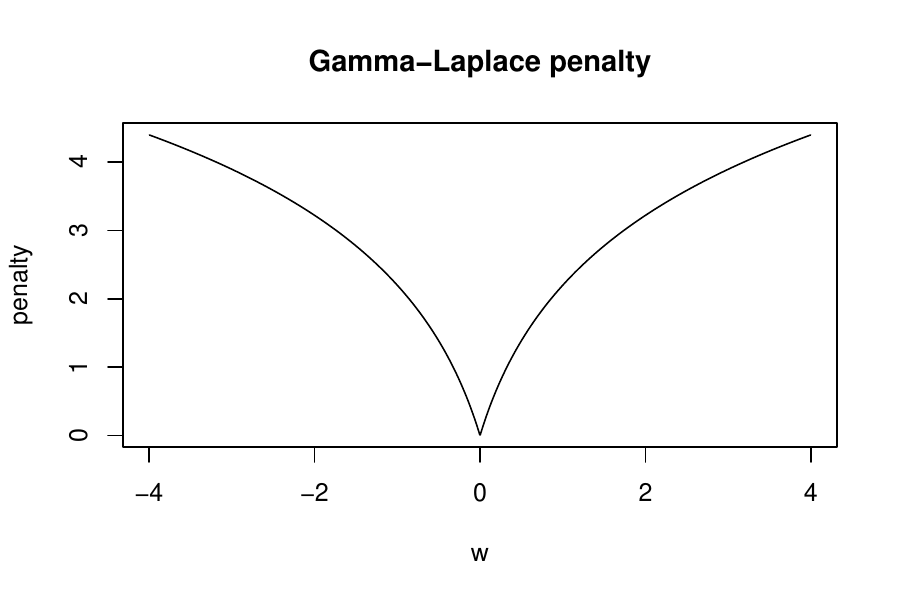}
    \vspace{-0.15in}
      \caption{Gamma-Laplace penalty $(s+D)\log(1+\sum_{d=1}^D|w_{d}|/r)$ for {$s=1$ and $r=1/2$.}}\label{fig:2.1}
\end{figure}

Unfortunately, available cross-validation (e.g., via solution paths) and fully Bayesian (i.e., through Monte-Carlo marginalization) methods for estimating $\vecw_{mg}$ under unknown $\lambda_{mg}$ are prohibitively expensive. Therefore, a novel algorithm is proposed for finding posterior mode estimates of the slope parameters, i.e., maximum \textit{a~posteriori} (MAP) estimation, while treating $\lambda_{mg}$ as missing data via an EM algorithm. The MAP inference with fixed $\lambda_{mg}$ is equivalent to likelihood maximization under an $L_1$-penalty in the LASSO  and $\lambda_{mg}\sim \text{Gamma} (s,r)$ leads to a non-convex penalty (Fig.~\ref{fig:2.1}):
\begin{equation*}\label{eq:2.5}
\begin{split}
C(\vecw_{mg})&=-\log\left\{\int_{\lambda_{mg}}\pi(\vecw_{mg}, \lambda_{mg}; s,r)d\lambda_{mg}\right\}\\
&=(s+D)\log\left(1+\sum_{d=1}^D|w_{dmg}|/r\right)+\text{constant},
\end{split}
\end{equation*}
for $s,r,\lambda_{mg}>0$.

By imposing different levels of regularization for $\vecw_g$, we consider two models herein. One model assumes component-and-variable-specific regularization, i.e., $\bs\lambda_g=\lambda_{1g},\ldots, \lambda_{mg}$; we call this model the general model. The second model assumes only component-specific regularization $\lambda_g$; we call this model the constrained model.  The penalty function for the constrained model is 
\begin{equation*}
\begin{split}
C(\vecw_{g})&=-\log \int_{\lambda_{g}}\pi(\vecw_{g}, \lambda_{g}; s,r)d\lambda_{g}\\
&=(s+D\times M)\log\left(1+\sum_{m=1}^{M}\sum_{d=1}^D|w_{dmg}|/r\right)+\text{constant},
\end{split}
\end{equation*}
for $s,r>0$. 
The constrained model can be useful when the number of latent traits $D$ is small.

\subsection{Motivation for Gamma-Laplace Penalties}\label{sec:GLpenalty}
One unique aspect of our approach is the use of independent gamma-Laplace priors for each slope parameter $\vecw_{mg}$. The Laplace prior for $\vecw_{mg}$ encourages sparsity in $\vecw_{mg}$ through a sharp density spike at $\vecw_{mg}=\mathbf{0}$, and MAP inference with fixed $\lambda_{mg}$ is equivalent to the likelihood maximization under an $L_1$ penalty in the LASSO estimation and selection procedure of \citet{tibshirani96}. In the Bayesian inference for LASSO regression, conjugate gamma hyerpriors are a common choice for the rate parameter $\lambda$ \citep[e.g.,][]{park08, yuan14}. However, the independent rate parameter $\lambda_{mg}$ is thought to provide a better representation of prior utility, and it avoids the over-penalization that can occur when inferring a shared rate parameter on clustered data.  

As detailed in Section~\ref{sec:nonconvexpenalties}, our approach yields an estimation procedure that corresponds to likelihood maximization under a specific non-convex penalty that can be seen as a re-parametrization of the ``log-penalty'' described in \citet{mazumder12}. Similar to the standard LASSO, singularity at zero in $C(\vecw_{mg})$ causes some coefficients to be set to zero. However, unlike the LASSO, the gamma-Laplace has gradient $$C'(\vecw_{mg})=\pm\frac{s+D}{\log\big(1+\sum_{d=1}^D|w'_{dmg}|/r\big)},$$ which disappears as $\sum_{d=1}^D|w_{dmg}|\rightarrow \infty$, leading to the property of unbiasedness for large coefficients \citep{fan01}.
Commonly, the rate parameter $\lambda$ is selected using cross-validation or an information criterion such as the BIC. %Bayesian information criterion \citep[BIC;][]{schwarz78}. 
However, our independent $\lambda_{mg}$ would require searches of impossibly massive dimension. Moreover, cross-validation is an estimation technique that is sensitive to the data sample on which it is applied. That said, one might wish to use cross-validation to choose $s$ or $r$ in the hyperprior and, because the results are less sensitive to these parameters than to a fixed penalty, a small grid of search locations should suffice. 

\subsection{Interpretation of the Model Parameters}\label{ch6:sec:interpretation}
The model parameters can be interpreted exactly as for the MLTA and item response models. In the finite mixture model, $\eta_g$ is the proportion of observations in the $g$th component. The characteristics of component~$g$ are determined by the parameters $\alpha_{mg}$ and $\vecw_{mg}$. In particular, the intercept $\alpha_{mg}$ has a direct effect on the probability of a positive response to the variable $m$ given by an individual in group $g$, through the relationship
\begin{equation*}
\pi_{mg}(0)=p(x_{im}=1|\vecy_i=0, z_{ig}=1)=\frac{1}{1+\exp(-\alpha_{mg})}.
\end{equation*}
The value $\pi_{mg}(0)$ is the probability that the median individual in group $g$ has a positive response for the variable $m$. However, when the dataset has a very low percentage of positive responses (e.g., text data), the value of $\pi_{mg}(0)$ can be very low for all items across all components. Thus, we use the slope parameters to characterize each component in Section~\ref{ch6:subsec:Reviews}. 

The slope parameters $\vecw_{mg}$ are known as discrimination parameters in item response theory. The larger the value of $w_{dmg}$, the greater the effect of factor $\vecy_d$ on the probability of a positive response to item $m$ in group~$g$.
The quantity $w_{dmg}$ can be used to calculate the correlation coefficient between the observed item $\vecx_{i}$ and the multivariate latent variable $\vecY_{i}$. In the latent trait case, the slope parameters cannot be interpreted as correlation coefficients because they are not bounded by $0$ and $1$. %as a correlation would be. 
However, it is possible to transform the loadings so that they can be interpreted as correlation coefficients in exactly the same way as in factor analysis. The standardized $w_{dmg}$ is given by 
\begin{equation*} \label{ch6:eq:stw}
w^*_{dmg}=\frac{w_{dmg}}{\sqrt{1+\sum_{d=1}^Dw_{dmg}^2}}.
\end{equation*}
The purpose of the Laplace prior for $\vecw_{mg}$ is to encourage sparsity in $\vecw_{mg}$, therefore identifying non-informative variables for each component. When the $m$th row of the slope parameter matrix for the $g$th component is zero everywhere ($w_{1mg}= w_{2mg}=\cdots =w_{dmg}=0$), then the corresponding variable is not informative. In addition, $w^*_{dmg}=0$ indicates that item $m$ is independent from latent trait $\vecy_d$ in component $g$.

\subsection{Model Identifiability}
The identifiability of our model depends on the identifiability of the latent trait part as well as the identifiability of the mixture model. The identifiability of finite mixture models has been discussed extensively \citep[e.g.,][]{mclachlan00}. \citet{knott99} discuss model identifiabilityin latent trait analysis. One necessary condition for model identifiability is that the number of free parameters to be estimated not exceed the number of possible data patterns. However, this condition is not sufficient because the actual information in a dataset can be less depending of the size of the dataset. As with the mixture of factor analyzers model, the slope parameters $\vecw_{mg}$ are only identifiable with $d\times d $ constraints. In the PMLTM, the model rotates the slope parameters automatically by shrinking the slope parameter through a penalized likelihood method. An empirical method can be used to assess non-identifiability; specifically, one can check whether the same maximized likelihood value is reached with different estimates of the model parameter values when starting the EM algorithm from different values. However, in Section~\ref{sec:simulation}, we showed that the standard deviations of $\vecw_{mg}$ estimates are relatively small from the 100 runs.

\section{Parameter Estimation and Implementation}
\subsection{Variational Approximation}\label{sec:variational}
\citet{jaakkola00} introduce a variational approximation for the predictive likelihood in a Bayesian logistic regression model and also briefly consider the ``dual'' problem, which is closely related to the latent trait model. Their method obtains a closed form approximation of the posterior distribution of the parameters within a Bayesian framework, and is based on a lower bound variational approximation of the logistic function
\begin{equation*}  
p(\vecx_{im}=1|\vecy_{i}, \, z_{ig}=1)=\frac{1}{1+\exp\{-(\alpha_{mg}+\vecw'_{mg}\vecy_{i})\}}.
\end{equation*}
This can be approximated by the exponential of a quadratic form involving variational parameters $\boldsymbol{\xi}_{ig}=(\xi_{i1g}, ... , \xi_{iMg})$, where $\xi_{img}\neq 0$ for all $m=1,...,M$. Now, the lower bound of each term in the log-likelihood is given by
{\small\begin{equation}  \label{eq:2.6}
L(\boldsymbol{\xi}_{ig})=\log\tilde{p}(\vecx_i|\boldsymbol{\xi}_{ig})=\log\left\{\int \prod_{m=1}^M \tilde{p}(x_{im}|\vecy_{i}, z_{ig}=1, \xi_{img}) p(\vecy_{i})\, d\vecy_{i}\right\},
\end{equation}}
where
{\small
\begin{equation*}\begin{split}  \label{eq:2.7}
&\tilde{p}(x_{im}|\vecy_{i}, z_{ig}=1,\xi_{img})=\sigma(\xi_{img})\exp\left\{\frac{A_{img}-\xi_{img}}{2}+ \lambda(\xi_{img})(A_{img}^2-\xi_{img}^2)\right\},\\
&A_{img}=(2x_{im}-1)(\vecw'_{mg}\vecy_{i}),\
\lambda(\xi_{img})=\frac{1}{2\xi_{img}}\left\{\frac{1}{2}-\sigma(\xi_{img})\right\},\\ 
&\sigma(\xi_{img})=\{1+\exp(-\xi_{img})\}^{-1}.
\end{split}\end{equation*}}
This approximation has the property that $\tilde{p}(x_{im}|\vecy_{i}, z_{ig}=1,\xi_{img})\le p(x_{im}|\vecy_{i}, z_{ig}=1,\xi_{img})$ with equality when $|\xi_{img}|=A_{img}$.

At first glance, the derivation of slope parameters $\vecw_{mg}$  is a challenging task due to the fact that the penalization term is not differentiable at $\vecw_{mg}$. To this end, we write $$|\vecw_{mg}|=\text{diag}\left(\sqrt{\vecw_{mg}\vecw_{mg}^{'}}\right)$$ and exploit the concavity of this square root. In particular, 
\begin{equation*}
-||\vecw_{mg}||_1\ge-\frac{1}{2}\left(\sum_{d=1}^D \frac{w_{dmg}^2}{|w_{dmg}w'_{dmg}|}+\sum_{d=1}^D|w_{dmg}^{*}| \right),
\end{equation*}
with equality if and only if $\vecw_{mg}=\vecw^{*}_{mg}$. Using these inequalities, we can obtain a surrogate function that can be used to obtain parameter estimates (Section~\ref{sec:par_est}).

\subsection{VEM Algorithm}\label{sec:par_est}
\subsubsection{Prior Specification} 
A classical assumption is to suppose independence among the prior distributions of the model parameters, thus,
\begin{equation*}
p(\bs \Theta)=\prod _{g=1}^Gp(\eta_g)\left\{\prod_{m=1}^M \prod_{d=1}^{D}p(w_{dmg}) p(\alpha_{mg})\prod_{i=1}^np(\xi_{img})\right\},
\end{equation*}
where $\eta_g \sim \text{Dirichlet}({1}/{2}, \ldots, {1}/{2})$, $\alpha_{mg} \sim N(0, 1)$, $ w_{dmg} \sim \text{Laplace}(0, \lambda_{mg})$, and $\xi_{img} \sim \text{uniform}[0,20]$.

\subsubsection {Parameter Estimation}
We use a VEM  algorithm to fit our model, which is a natural approach for MAP estimation when data are incomplete. Each iteration of the VEM algorithm has two steps: a variational expectation (VE) step, where we approximate the logarithm of the component densities with a lower bound, and a maximization (M) step, where the log (complete-data) posterior is maximized with respect to the model parameters. Our case features three sources of missing data: $\{\vecz_{i}\}_{i=1}^n$ arises from the fact that we do not know the cluster labels, $\{\vecy_{i}\}_{i=1}^n$ are realizations of the $D$-dimensional continuous latent variable, and $\{\bs \lambda_{m}\}_{m=1}^M$ are the unknown Laplace rate parameters. 
The purpose of the M-step of the VEM algorithm is to find the MAP estimates of $\bs \Theta$ by maximizing the conditional expectation of the log (complete-data) posterior %$\log p(\bs \Theta|\bs x, \bs y, \bs z, \bs \lambda)$ which can be easily obtained:
%\begin{equation*}\label{eq:2.7}
%Q(\bs \Theta_g|\bs \Theta_g^{(t)}) %\log p(\bs \Theta|\vecx, \vecy, \vecz, \bs \lambda) 
%= \sum_{i=1}^n\sum_{g=1}^G\left\{\mathbb{E}_{p(z_{ig}\mid\vecx_i, \bs \theta_g)}\left[\log p(\vecx_i, z_{ig} |\bs \theta_g, \vecy_{i})\right]\right\}+\sum_{g=1}^G\left(\log p(\vecw_g,\bs \lambda_{g})+\mathbb{E}_{p(\bs \lambda_g\mid \vecw_g)}\right).
%\end{equation*}
%\begin{equation}\label{eq:2.7}
%\log p(\bs \Theta|\vecx, \vecy, \vecz, \bs \lambda) = \sum_{i=1}^n \sum_{g=1}^G \log p(\vecx_i|\bs \theta_g, \vecy_{i}, z_{ig})+\log p(\vecw_g|\bs \lambda_{g}).
%\end{equation}
\begin{equation*}\label{eq:2.7}
\begin{split}
Q(\bs \Theta_g|\bs \Theta_g^{(t)}) %\log p(\bs \Theta|\vecx, \vecy, \vecz, \bs \lambda) 
&= \sum_{i=1}^n\sum_{g=1}^Gz_{ig}\big\{\log\eta_g+\log p(\vecx_i |\bs \theta_g, \vecy_{i}, z_{ig}=1)+\log p(\vecy_i)\big\}
\\&+\sum_{g=1}^G\log p(\vecw_g,\bs \lambda_{g}).
\end{split}
\end{equation*}

\textbf{VE-step.}
We compute the expected value of the complete-data log-posterior in the VE-step using the expected values of the missing data in $\log p(\bs \Theta|\vecx, \vecy, \vecz, \bs \lambda)$. We require the expectations
\begin{equation*}
	z_{ig}^{(t+1)}=\frac{\eta_g^{(t)} \exp\{L(\bs{\xi}_{ig}^{(t)})\}}{\sum_{g=1}^G \eta{'}_{g}^{(t)} \exp\{L(\bs{\xi}{'}_{ig}^{(t)})\}}.
\end{equation*}

Compute the location vector and the covariance matrix for \\$\tilde{p}(\vecy_i|\vecx_i, z_{ig}^{(t+1)}, \bs \xi_{ig}^{(t)}, \bs \alpha_{g}^{(t)}, \vecw_{mg}^{(t)})$, which is an $\mvn(\bs \mu_{ig}^{(t+1)}, \bs \Sigma_{ig}^{(t+1)})$ density with
\begin{equation*}
\begin{split}
&\bs \Sigma_{ig}^{(t+1)}=\left\{\mathbf{I}_D-2\sum_{m=1}^M B(\xi_{img}^{(t)}) \vecw_{mg}^{(t)} \vecw_{mg}^{'(t)} \right\}^{-1},\\
&\bs \mu_{ig}^{(t+1)}=\bs \Sigma_{ig}^{(t+1)}\left[\sum_{m=1}^M\left\{x_{im}-\frac{1}{2}+2B(\xi_{img}^{(t)})\alpha_{mg}^{(t)}\right\}\vecw_{mg}^{(t)},\right],
\end{split}
\end{equation*}
where $$B(\xi_{img}^{(t)})=\frac{{1}/{2}-\sigma(\xi_{img}^{(t)})}{2\xi_{img}^{(t)}} \quad \text{ and }\quad \sigma(\xi_{img}^{(t)})=\frac{1}{1+\exp\big\{-\xi_{img}^{(t)}\big\}}.$$
The expected value of the independent multidimensional rate parameter can be written
\begin{equation*}
\lambda_{mg}^{(t+1)}=\frac{s+D}{\sum_{d=1}^D|w_{dmg}^{(t)}|+r},
\end{equation*}
where $s, r>0$ are predetermined shape and rate parameters of the gamma hyperprior, respectively.\\[-6pt]

\textbf{M-step 1.} The first M-step on the $(t+1)$ iteration optimizes the variational parameter $\xi_{img}$ to make the approximation $\tilde{p}(\vecx_i|z^{(t+1)}_{ig}=1,\bs\xi^{(t+1)}_{ig})$ as close as possible to $p(\vecx_i|z_{ig}=1)$:
\begin{equation*}
\big(\xi_{img}^2\big)^{(t+1)}=\vecw_{mg}^{'(t)}\left(\bs \Sigma_{ig}^{(t+1)}+\bs \mu_{ig}^{(t+1)} \bs \mu_{ig}^{'(t+1)}\right)\vecw_{mg}^{(t)}+2\alpha_{mg}^{(t)}\big(\vecw_{mg}^{(t)}\big)' \bs \mu_{ig}^{(t+1)}+\big(\alpha_{mg}^{(t)}\big)^2.
\end{equation*}

\textbf{M-step 2.} The second M-step optimizes the parameters $\vecw_{mg}$ and $\bs \alpha_{g}$ to increase the log (complete-data) posterior $$\log p(\vecw_{mg}, \bs \alpha_g|\vecx, \vecy^{(t+1)},\bs \lambda_g^{(t+1)},\bs \xi_g^{(t+1)}, \vecz_g^{(t+1)}),$$ where $\vecw_{mg}$ and $\bs \alpha_g$ are given in Appendix~\ref{sec:appmath} along with a more convenient form of the update for $\vecw_{mg}$.\\[-6pt]

\textbf{M-step 3.} The third M-step updates $\eta_g$ via
\begin{equation*}
\eta_g^{(t+1)}=\frac{n_g^{(t+1)}}{n},
\end{equation*}
where $n_g^{(t+1)}=\sum_{i=1}^nz_{ig}^{(t+1)}$.\\[-6pt]

\textbf{Compute the log-likelihood.} Obtain the lower bound of the log-likelihood at the expansion point $\xi_{ig}$:
{\small
\begin{equation*}
\begin{split}
L(\boldsymbol{\xi}_{ig}^{(t+1)})=&
\sum_{m=1}^M\Bigg\{\log\sigma(\xi_{img}^{(t+1)})-\frac{\xi_{img}^{(t+1)}}{2}-B(\xi_{img}^{(t+1)})\big(\xi_{img}^{(t+1)}\big)^2+\left(x_{im}-\frac{1}{2}\right)\alpha_{mg}^{(t+1)}\\& \quad+B(\xi_{img}^{(t+1)})\big(\alpha_{mg}^{(t+1)}\big)^2\Bigg\}+\log\frac{|\boldsymbol{\Sigma}_{ig}^{(t+1)}|}{2}+\frac{\bs \mu_{ig}^{'(t+1)} [\bs\Sigma_{ig}^{(t+1)}]^{-1} \bs \mu_{ig}^{(t+1)}}{2}
\end{split}
\end{equation*}}
and the log-posterior:
\begin{equation*}
l^{(t+1)}\approx\sum_{i=1}^n\log\left[\sum_{g=1}^G\eta_g^{(t+1)}\exp\{L(\boldsymbol{\xi}_{ig}^{(t+1)})\}\right]+\sum_{g=1}^G\log p(\vecw_g^{(t+1)}).
\end{equation*}

\textbf{Convergence criterion.} The convergence of the variational EM algorithm is determined using a criterion based on the Aitken acceleration \citep{aitken26}, i.e.,
$|l_{\infty}^{(t+1)}-l_{\infty}^{(t)}|<0.01,$
where 
$$l_{\infty}^{(t)}=l^{(t-1)}+\frac{1}{1-a^{(t-1)}}(l^{(t)}-l^{(t-1)}), \qquad
a^{(t)}=\frac{l^{(t+1)}-l^{(t)}}{l^{(t)}-l^{(t-1)}},$$
and $l^{(t)}$ is the log posterior at iteration $t$ \citep{bohning94}.

\subsection {Model Selection}\label{sec:modelselection}
The BIC \citep[][]{schwarz78} is used a criterion for model selection:
\begin{equation}\label{eqn:bic}
\text{BIC}=-2l+\text{k}\log n,
\end{equation}
where $l$ is the maximized log-likelihood, k is the number of free parameters to be estimated in the model, and $n$ is the number of observations. The presence of a penalty term reduces the number of free parameters of the slope parameter matrix because the effective degrees of freedom of $\vecw$ equals the number of nonzero terms in the loading parameter matrix. 

Within the framework of MLTA models, the number of components $G$ and the dimension of the latent variable $\vecY$ (i.e., $d$) need to be determined. When defined as in \eqref{eqn:bic}, models with lower BIC values are preferred. The BIC value could be overestimated using the variational approximation of the log-likelihood, which is always less than or equal to the true value. For model selection purposes, we calculate the maximum log posterior using Gauss-Hermite quadrature after convergence is attained. In Section~\ref{sec:sim3}, we demonstrate that the BIC is effective for choosing the correct number of components $G$ and the dimension of the latent variable $\vecY$.

For high-dimensional binary data, particularly when the number of observations $n$ is not very large relative to their dimension $m$, having a large number of patterns with small observed frequency is common. Accordingly, we cannot use a $\chi^2$ test to check the goodness of the model fit. In the simulated examples in Section~\ref{sec:simulation}, where the true classes are known, the adjusted Rand index \citep[ARI;][]{hubert85} can be used to assess model performance. The ARI is the corrected-for-chance version of the Rand index \citep{rand71}. The %general form of the 
ARI %is $$\frac{\text{index}-\text{expected index}}{\text{maximum index}-\text{expected index}},$$ which 
is bounded above by 1, and has expected value 0 under random classification. %An ARI value of~1 corresponds to perfect agreement, and a value of 0 would be expected under random classification. 
In real examples, such as the analysis of the Boston Airbnb reviews in Section~\ref{ch6:subsec:Reviews}, the analysis of the clusters in the selected model can be used to interpret the model.
 
\subsection{Selection of Programming Languages}
When fitting the PMLTM model using {\sf R}, the task becomes increasing burdensome as the number of items becomes large. Therefore, we implement our algorithm in two scripting languages in Section~\ref{sec:sim1}, {\sf R} and Python, and compare their performance (Table~\ref{table:time}, Appendix~\ref{app:addtables}). Python is an elegant open-source language that has become popular in the scientific community. We use the \texttt{Numpy} library for matrix operations and \texttt{Scipy.stats} library for probability distributions and statistical functions. %Moreover, Python does not generate copies of the objects in an array or a list when slicing arrays and lists which may save memory and shorten the runtime. 

\subsection{Simulation Studies} \label{sec:simulation}
\subsubsection{Overview}
Simulation studies are performed to illustrate the proposed PMLTM model with data generated in a number of scenarios, resulting in four experiments (see Table \ref{SimDetail} for details). A set of 100 samples for each scenario is generated from an MLTA model with a $G$-component mixture with $\pi_1=\pi_2\cdots=\pi_G={1}/{G}$. The latent variable is generated from a $D$-variate Gaussian distribution, i.e., $\vecY \sim \mvn(\mathbf{0}, \mathbf{I})$, and two out of $D$ elements of $W_{mg}$ are randomly set to zero. We only fit the general model to these experiments.
\begin{table}
\caption{\label{SimDetail}Key characteristics for Experiments~1--4.}
	\centering
\begin{tabular}{ccccc}
	\hline
Experiment&$n$&$G$&$M$&$D$\\
\hline
1&$\{300, 600, 900\}$&$\{2\}$&20&\{3\}\\
2&$\{300\}$&$\{2\}$&20&\{10\}\\	
3 &$\{300, 600, 900\}$&$\{2,3,5\}$&20&\{3,5,10\} \\
4 &$\{300, 600, 900\}$&$\{2,3,5\}$&20&5\\
\hline 
\end{tabular}
\end{table}

We assess the performance of the PMLTM in a few different ways: Experiment 1 (Section~\ref{sec:sim1}) illustrates the ability of our proposed model to recover underlying parameters when the number of components and the model are correctly specified; then we investigate the model sensitivity to different values of the gamma hyperparameters, i.e., $(s,r)$, in Section~\ref{sec:sim2} (Experiment 2). We study the effectiveness of the BIC for choosing the correct model in Experiment 3 (Section~\ref{sec:sim3}), and last but not least, we compare our proposed model with the MLTA and LDA in Section~\ref{sec:sim4} (Experiment 4).

\subsubsection{Parameter Recovery under the True Model} \label{sec:sim1}
The first experiment is designed to evaluate the ability of our proposed model to recover underlying parameters when the number of components and latent traits are correctly specified. The means of the parameter estimates and their associated standard deviations are summarized in Table~\ref{sim1result} (Appendix~\ref{app:addtables}). The results show that the means of all parameter estimates are close to each other and the standard deviations decrease when the number of observations increases.

\subsubsection{Model Sensitivity to Values of the Gamma Hyperparameters} \label{sec:sim2}
In Experiment 2, the values of the gamma hyperparameters, i.e., $(s,r)$, are selected from $\{(0.1,0.5), (0.5,0.5), (1,0.5), (2,0.5)\}$. Table~\ref{table:Simsr} shows the BIC and ARI values averaged over the 100 samples for each pair $(s, r)$.
The clustering results do not vary much for different values of the gamma hyperparameters.
We choose to use $(s,r)=(1,0.5)$ for the rest of the paper because, on average, the BIC has a minimum when $(s,r)=(1,0.5)$.
\begin{table}
\caption{\label{table:Simsr}BIC and ARI values averaged over the $100$ samples for each $(s,r)$. }
 \centering
 \begin{tabular}{ r c c c c  }
 \hline
 &$s=0.1,r=0.5$&$s=0.5,r=0.5$&$s=1,r=0.5$&$s=2,r=0.5$\\
 \hline
 \hspace{0.2cm}BIC&13389&13521&12234&12490\\
 ARI&0.70&0.72&0.72&0.72\\ 
 \hline
 \end{tabular}
 \end{table}%

\subsubsection{Effectiveness of the BIC for Choosing the Correct Model} \label{sec:sim3}
The PMLTM model is fitted to these data for $G=1,\ldots,6$ and $D=1,\ldots, 10$. Table~\ref{table:simBIC} summarizes the number of times that the correct model is favoured by the BIC for each $n$, $G$, and $D$ as well as the ARI averaged over the 100 replications. The BIC selects the correct model on most occasions and the ARI increases significantly when the number of observations reaches 900. Therefore, we are confident that the BIC is effective at choosing the number of components and the latent traits when the number of nonzero parameters is used as $k$.
\begin{table}
	\caption{\label{table:simBIC}Percent preferred by the BIC and average ARI (replications=100) with $G=1,\ldots,6$ and $D=1,\ldots, 10$.}	
     \centering		
		\begin{tabular}{rccccccc}
		  \hline
 		&&\multicolumn{2}{c}{$G=2$} & \multicolumn{2}{c}{$G=3$} & \multicolumn{2}{c}{$G=5$} \\
		\cline{3-4}\cline{5-6}\cline{7-8}
		               &&BIC&ARI&BIC&ARI&BIC&ARI\\
 		 \hline
\multirow{3}{*}{\hspace{0.2cm}$n=300$}&$D=3$&94&0.69&92&0.68&88&0.68\\
&$D=5$&92&0.69&92&0.70&90&0.70\\
&$D=10$&92&0.72&91&0.72&90&0.72\\
\hline
\multirow{3}{*}{$n=600$}&$D=3$&95&0.78&90&0.76&89&0.74\\
&$D=5$&99&0.80&97&0.76&95&0.74\\
&$D=10$&96&0.80&95&0.78&90&0.76\\
\hline
\multirow{3}{*}{$n=900$}&$D=3$&100&0.90&100&0.85&95&0.85\\
&$D=5$&100&0.89&100&0.85&100&0.83\\
&$D=10$&95&0.90&96&0.80&95&0.88\\
\hline
	\end{tabular}
\end{table}

\subsubsection{Comparison with the MLTA and LDA} \label{sec:sim4}
Finally, we compare our proposed model with the MLTA and LDA (via {\sf R} package \texttt{topicmodels}). The LDA approach is fitted with the correct number of components (topics), and we assign an observation to a specific cluster when a topic has the highest contribution. The ARI* is the average ARI when the correct number of components is selected (see Table~\ref{table:compare}, Appendix~\ref{sec:sim4k}). Not surprisingly, the LDA approach does not perform well using data generated from our model. MLTA and PMLTM both yield excellent clustering results when $G=2$ and $G=3$. The performance of the MLTA drops significantly for $G=5$ while that of the PMLTM remains the same. 
%\begin{table}
%	\caption{\label{table:compare}Percent preferred by the BIC and average ARI with $G=1, \ldots, 6$.}	
%	\centering
%	\fbox{%
%		\begin{tabular}{rccccccc}
%		  \hline
% 		&&\multicolumn{2}{c}{PMLTM} & \multicolumn{2}{c}{MLTA} & \multicolumn{2}{c}{LDA} \\ 
%				\cline{3-4}\cline{5-6}\cline{7-8}
%		               &&BIC&ARI*&BIC&ARI*&BIC&ARI\\
% 		 \hline
%\multirow{3}{*}{\hspace{0.2cm}$n=300$}&$G=2$&92&0.74&92&0.72&N/A&0.10\\
%&$G=3$&92&0.72&85&0.70&N/A&0.15\\
%&$G=5$&90&0.72&67&0.24&N/A&0.18\\
%\hline
%\multirow{3}{*}{$n=600$}&$G=2$&99&0.85&94&0.86&N/A&0.26\\
%&$G=3$&97&0.83&89&0.84&N/A&0.20\\
%&$G=5$&95&0.80&86&0.57&N/A&0.18\\
%\hline
%\multirow{3}{*}{$n=900$}&$G=2$&100&0.89&96&0.90&N/A&0.29\\
%&$G=3$&100&0.85&94&0.85&N/A&0.30\\
%&$G=5$&100&0.83&90&0.57&N/A&0.24\\
%\hline
%	\end{tabular}}	
%\end{table}%

\section{Boston Airbnb Reviews} \label{ch6:subsec:Reviews}
\subsection{Data and preprocessing}
\begin{figure}[htb]
        \centering
      \includegraphics[width=0.85\textwidth]{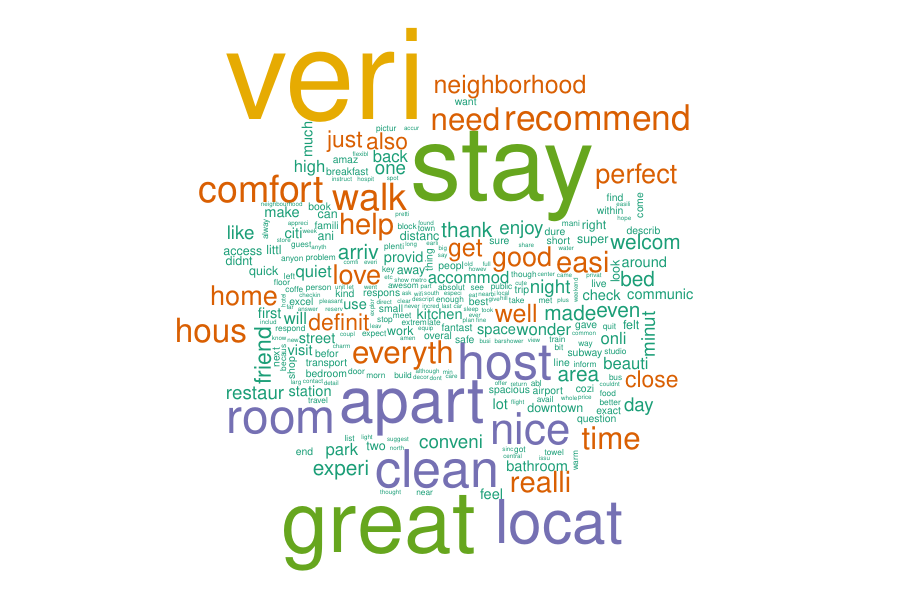}
      \caption{Word cloud for the Airbnb comments.}\label{ch6:fig:wordcloud}
\end{figure}

From March 2009 to September 2016, there are detailed English comments for 2,829 hosts on the Airbnb website in the Boston area from 63,812 guests  (i.e., $n=63,812$). To get meaningful guest segmentation, we only take the most recent comment from a guest if they left more than more comments within the time period. This dataset is available from {\tt kaggle.com}. Comments with non-ASCII (American standard code for information interchange) characters are removed to ensure we only include English comments. We perform some pre-processing of the text data (i.e., converting the text to lower case, removing numbers and punctuation, removing stop words, and stemming). These basic transforms are available within the {\sf R} package \texttt{tm} \citep{feinerer15}. The term matrix with each comment as a row and each word as a column is then created. If a word is mentioned in a comment, the response for the corresponding cell is coded as 1, and otherwise is 0. The term matrix contains 43,584 words but most are infrequently used, i.e., so-called ``sparse terms''. Sparse terms that appear in less than 1\% of all reviews are not of interest and so are removed. %because we are often not interested in such terms. 
At the end of this pre-processing step, the term matrix consists of $473$ words (i.e., $M=473$) and the word cloud in Fig.~\ref{ch6:fig:wordcloud}  provides a quick visual overview of the frequency of the words in the final term matrix.

\subsection{Topic Findings via LDA}
\begin{figure}
	\centering
	\includegraphics[width=0.7\textwidth]{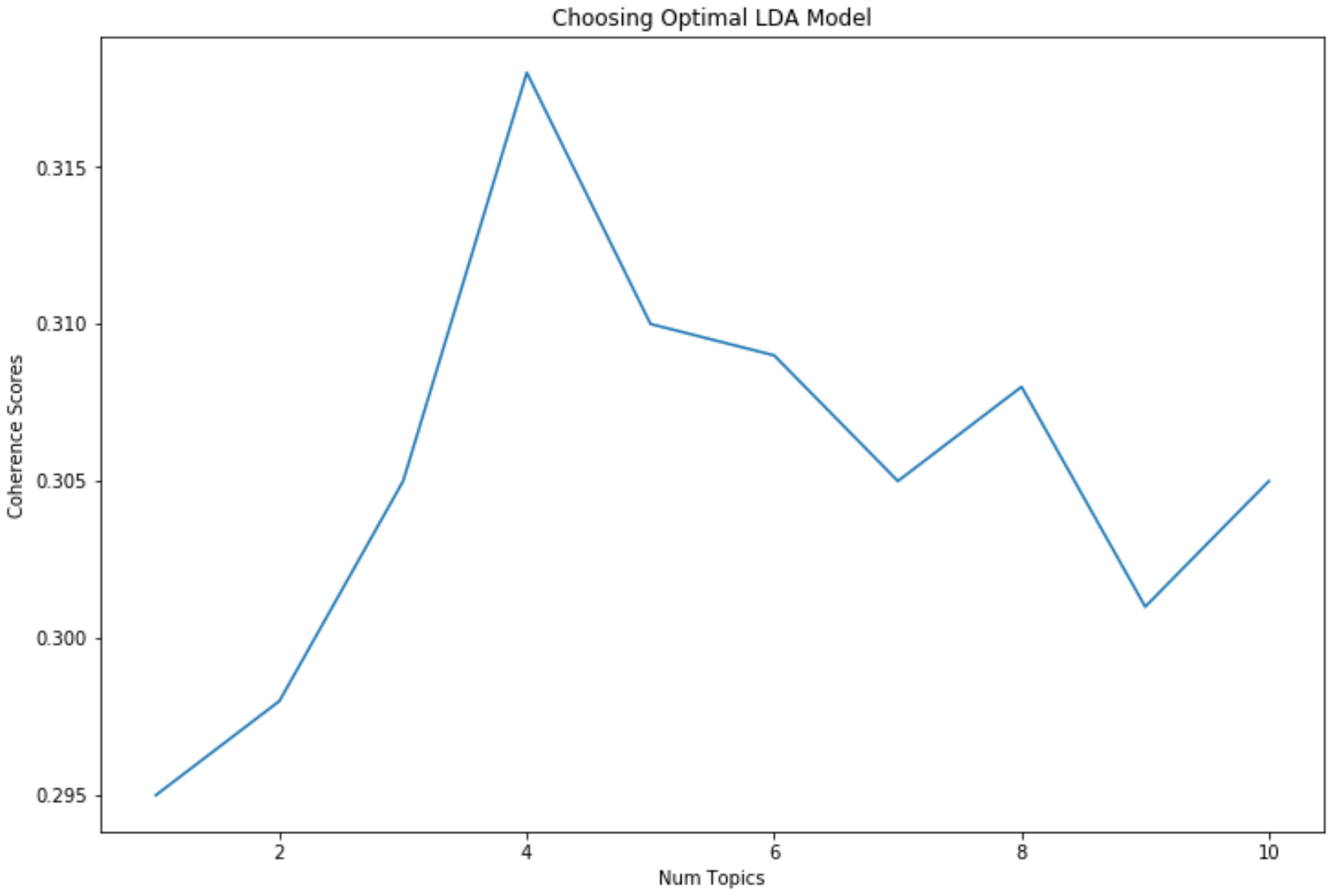}
	\vspace{-0.2in}
	\caption{Averaged coherence scores vs.\ the number of topics using LDA.}\label{fig:ldatopics}
\end{figure}

LDA is one of the most popular unsupervised learning techniques for finding topics in documents. We fit the data for $k$ topics, $k=1, \ldots, 10$, and then we find the optimal number of topics by picking the one that gives the highest coherence value. Figure~\ref{fig:ldatopics} shows the coherence score increases with the number of topics. 

Four topics have been discovered within the reviews using LDA including location (Topic 3), amenities (Topic 2), host (Topic 1) and recommendation (Topic 4). Table~\ref{table:ldawords} shows the top ten keywords for each of these four topics. The keywords associated with `host' describes the helpfulness of the host as well as the home feeling the guests experienced with Airbnb. The topic `amenities' covers the room environment such as bed, kitchen and bathroom. Several location related keywords such as walk, subway, and restaurant can be found in the third topic and Topic~4 leads to a recommendation of a place.  
\begin{table}
  \caption{  \label{table:ldawords} Top ten keywords for each topic (Tp.) using LDA for the Airbnb data.}
  \centering
 \begin{tabular}{lp{5in}}
\hline
Tp. & Top 10 keywords\\
\hline
1& stay, host, home, make,  place, welcom, great,  hous, comfort, help\\
2 &get, bed, night, apart, kitchen, bathroom , clean, stay, need,  use  \\
3& walk, restaur, station, minut, close, distanc, locat,  subway, shop, line    \\
4&locat, great, stay, apart, place, clean, would, recommend, host, everyth \\
\hline
 \end{tabular}%
\end{table}% 

\subsection{Sentiment Analysis Result}
\begin{figure}[htb]
        \centering
      \includegraphics[width=0.7\textwidth]{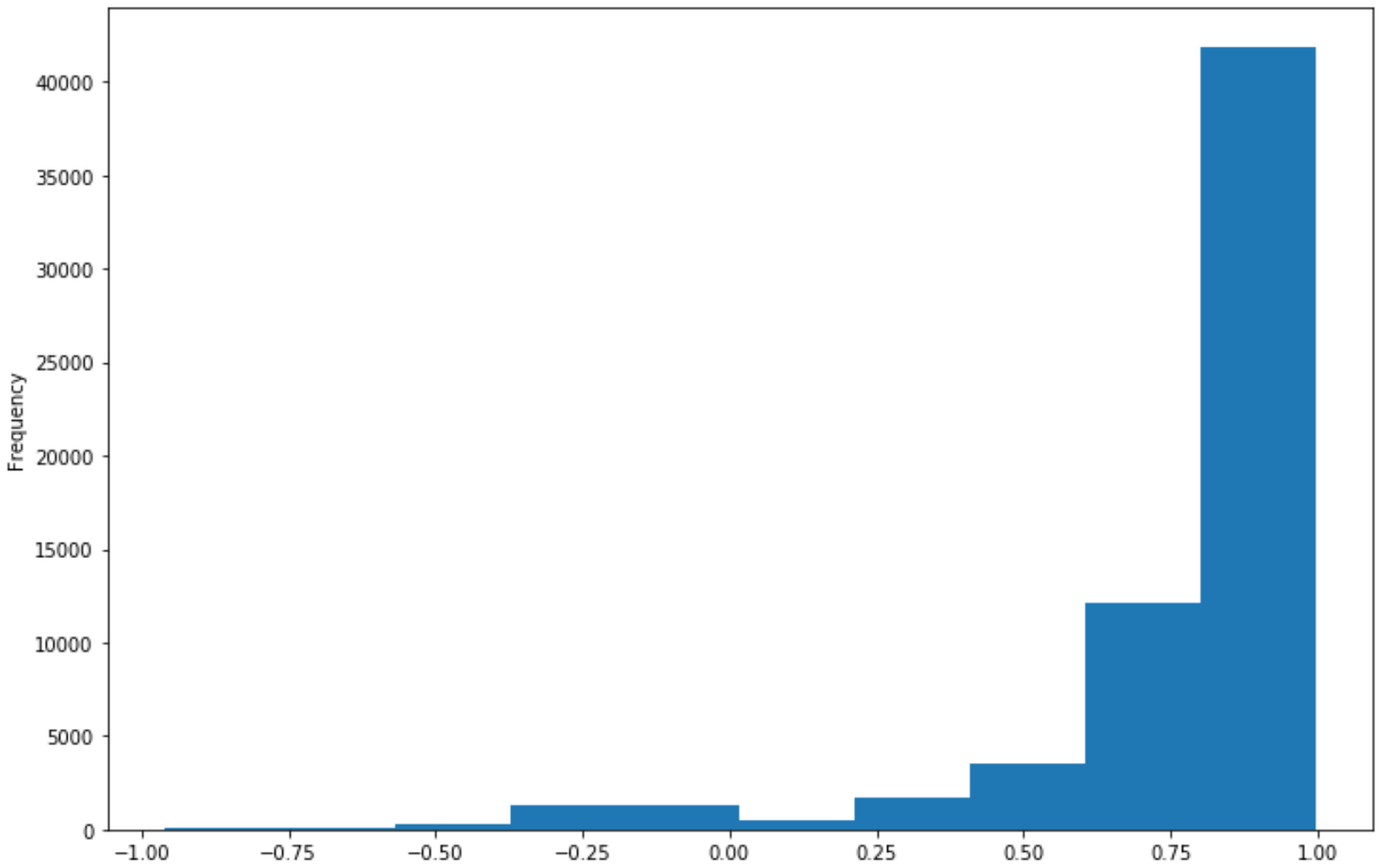}
      \vspace{-0.2in}
      \caption{A frequency histgram of compound scores for the Airbnb reviews. }\label{fig:compoundhist}
\end{figure}

\begin{figure}[htb]
        \centering
      \includegraphics[width=0.7\textwidth]{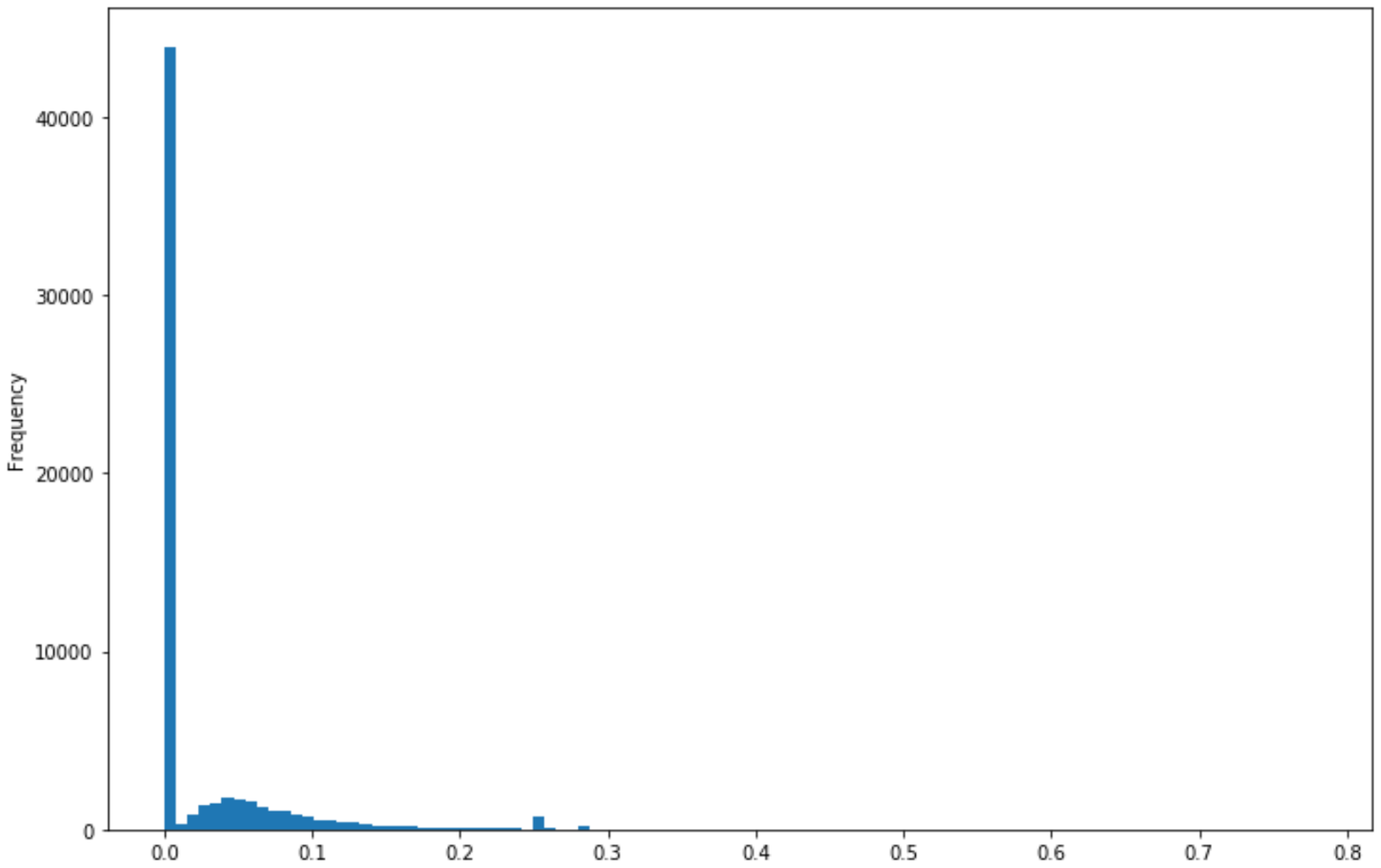}
      \vspace{-0.2in}
      \caption{A frequency histogram of negative scores for the Airbnb reviews. }\label{fig:neghist}
\end{figure}

Average sentiment scores are calculated using the built-in Python library \texttt{nlkt} \citep[natural language toolkit;][]{bird04}. The sentiment of each comment---positive, negative, or neutral---is sumarized using a score ranging from 0 to 1. Because each comment could contain positives and negatives at the same time, a compound score is presented as well. Each compound score is presented over $[-1, 1]$, where $-1$ corresponds to an overall unpleasant tone and $1$ is an overall pleasant tone. The result of the sentiment analysis indicates that Airbnb users are overwhelmingly positive about their experiences (Figure~\ref{fig:compoundhist}). There are only very few reviews are classified as having significant amounts of negativity. In addition, a significant amount of the reviews are given exactly $0.0$ negativity (Figure~\ref{fig:neghist}). 

\subsection{Partition Study}
The PMLTM model is fitted to these data for $D=1,\ldots,5$ and $G=1,\ldots,5$. We run all models using Python. The minimum BIC value (954,147) occurs for the four-component, two dimensional (i.e., two latent traits) constrained PMLTM model. The clusters (components) for the selected model $(G=4, d=2)$ are summarized in Table~\ref{ch6:table:4.4}. 
Average sentiment scores for each cluster are calculated. Cluster~1 consists of comments with an overall pleasant tone, as indicated by the compound score of $0.91$. The average positivity score in Cluster~1 is 0.28. Cluster~3 consists of positive-negative comments where negativity is presented in the reviews. Cluster~2 is a small group that consists of comments that have a slightly negative overall tone. Noting that the average negativity score is higher than the average positivity score in Cluster~2. The overall compound score of Cluster~4 is the highest among the four clusters (0.96), with an average positivity score of 0.35.
\begin{table}
	\caption{\label{ch6:table:4.4}The predicted classification and sentiment scores for the chosen PMLTM model ($G=4$, $D=2$ constrained) for the Airbnb data.}
	\centering
	\begin{tabular}{r c cccc }
	\hline 
	&No.\ Obs.&Compound&Negativity&Neutrality&Positivity \\ [0.5ex] % inserts table %heading
	\hline 
	\hspace{0.2cm}Cluster~1&28492&0.91&0.03&0.68&0.28\\
	Cluster~2&4382&-0.37&0.07&0.88&0.05\\
	Cluster~3&10782&0.60&0.08&0.72&0.20\\
	Cluster~4&20156&0.96&0&0.65&0.35\\
	\hline
	\end{tabular}
\end{table}

In addition, a logical approach to classify a review as belonging to one of the four topics is to see which topic has the highest contribution to that review and assign it. The clustering results are presented in Table~\ref{table:ldaresult}. At first glance, the two approaches yield very different results: LDA finds four evenly distributed clusters while PMLTM is able to find a small cluster (i.e., Cluster 2) consisting of comments that have a slightly negative overall tone. This is because LDA only finds topics while PMLTM takes into account both topics and sentiment. 
\begin{table}
\caption{\label{table:ldaresult}Predicted classification and sentiment scores for the LDA for the Airbnb data.}
\centering
\begin{tabular}{r ccccc }
\hline 
&No.\ Obs.&Compound&Negativity&Neutrality&Positivity \\ [0.5ex] % inserts table %heading
\hline 
\hspace{0.2cm}Cluster~1& 16114 &0.96&0&0.65&0.35\\ 
Cluster~2&15385& 0.86&0&0.63&0.37\\
Cluster~3&14892&0.93&0&0.68&0.32\\
Cluster~4&17421&0.71&0.05&0.89&0.08\\[1ex]
\hline
\end{tabular}
\end{table}
\subsection{Interpretation of the Best Model}
Table~\ref{ch6:table:4.6} shows the high-loading words for each latent trait in each cluster. We note that the first latent trait $\vecY_1$ is concerned with the property (e.g., location, condition, etc.) whereas the second latent trait $\vecY_2$ is concerned with the host. The two traits are consistent with the topics found using LDA where $\vecY_1$ represents `location' and `amenities',  and $\vecY_2$ represents `recommendation' and `host'.
\begin{table}
  \caption{\label{ch6:table:4.6}High-loading words for each latent trait of our chosen model ($G=4$, $d=2$) for the Airbnb data.}
  \centering
 \begin{tabular}{rlp{5in}}
\hline
\multirow{ 2}{*}{\hspace{0.2cm}Cluster~1}&$\bs Y_1$&{bar, bedroom, big, bus, easili, equip, experi, floor, lot, metro, minut, store}\\
&$\bs Y_2$&{answer, anything, apprici, ask, list, met, return, reserv}\\\hline
\multirow{2}{*}{Cluster~2}&$\bs Y_1$&{busi, discript, cute, detail, ever, par, never, old, explor, mattress, north,  south, studio}\\
&$\bs Y_2$&{checkin, common, contact, couldn't, disappoint, quit, suggust}\\\hline
\multirow{2}{*}{Cluster~3}&$\bs Y_1$&{found, stay, spot, care, valu, nothing}\\
&$\bs Y_2$&{welcome, help, pleasant, next, good, book, friend, suggest, though}\\ \hline
\multirow{ 2}{*}{Cluster~4}&$\bs Y_1$&{close, home, walk}\\
&$\bs Y_2$&{absolute, amaz, everyth, realli, easi, arriv}\\
\hline
    \end{tabular}%
 \end{table}% 

%We can further characterize our groups from this observation. 
Based on the characteristics of the clusters, the four clusters were named property driven guests (Cluster~1), host-driven guests (Cluster~4), guests with recent overall negative stay (Cluster~2) and guests with some negative experiences (Cluster~3). With PMLTM, not only we are able to identify groups of guests with different preferences (property driven vs. host driven), we are also able to find a small cluster that mainly consists of negative reviews simultaneously.  Moreover, the positive comments are more intense in Cluster~4 because words such as ``absolute'', ``amazing'' and ``everything'' are used. From the high-loading words for Cluster~3, we say guests from this groups were mainly positive with the host but negative with the property. Examples of comments from each cluster are given in Tables~\ref{ch6:table:4.5}--\ref{ch6:table:4.5c} (Appendix~\ref{sec:sim4k}). 

\section{Summary}\label{sec:discussion}
%We extend the MLTA model by introducing gamma-Laplace penalties on the slope parameters. The PMLTM model enables us to 
An MLTA approach that encourages sparsity in estimating the slope parameters --- thus considerably reducing the number of free parameters --- is introduced for clustering the Boston Airbnb data. The component-specific rate parameters avoid the over-penalization that can occur when inferring a shared rate parameter on clustered data. The PMLTM model retains the ability to investigate the dependence between variables while clustering with the added advantage of being able to model very high-dimensional binary data (e.g., text data).
%
%The excellent clustering behaviour of this method has been shown by two applications: on the U.S. Congressional Voting and the Boston Airbnb reviews datasets. In both cases, the model found groups that were intuitive in their interpretation. 
%
Applying the PMLTM model to the Boston Airbnb reviews data shows that the method scales to far larger datasets than any existing model-based clustering methods for binary data. The results for these data reveal two latent traits and four clusters of reviews. The latent traits can be interpreted as concerning the property and the host, respectively. One cluster contains highly positive property driven reviews, another contains highly positive host driven reviews, the third contains some negative regarding the property, and the last contains reviews that are not positive, i.e., moderate and negative reviews. 
%
%A parsimonious family of the mixture of latent trait models is developed by using common slope parameters and applying restrictions to the components of the decomposed covariance matrices in \citet{tang15}. Analogous families of parsimonious models could be developed for the PMLTM model to further reduce the number of parameters to be estimated, which would make this model even more powerful for the analysis of high-dimensional binary data. Probabilities on restricting the turning parameters $\bs \lambda$ can be explored as well. 
%
These results could be used in several ways. For example, hosts can use the high-loading words in a cluster to better understand what is driving consumer opinion. Understanding consumer subgroups can also be useful to Airbnb, who could provide specific (e.g., location-based) suggestions to hosts on how they should advertise their particular property.

{\small

\appendix

\section{Values of $\vecw_{mg}$ and $\bs \alpha_g$ for M-step 2}\label{sec:appmath}
For M-step 2, we have 
{\footnotesize\begin{equation*}
\begin{split}
&\vecw_{mg}^{(t+1)}=\left[2\sum_{i=1}^n z_{ig}^{(t+1)}B(\xi_{img}^{(t+1)})\mathbb{E}(\vecY_{ig}\vecY'_{ig}) \right. \\
&\left.+\frac{2\{\sum_{i=1}^n z_{ig}^{(t+1)} B(\xi_{img}^{(t+1)}) \mathbb{E}(\vecY_{ig})\}\{\sum_{i=1}^n z_{ig}^{(t+1)} B(\xi_{img}^{(t+1)}) \mathbb{E}(\vecY'_{ig})\}}{\sum_{i=1}^n z_{ig}^{(t+1)} B(\xi_{img}^{(t+1)})-n_g^{(t+1)}}-\lambda_{mg}\bs \Gamma_{mg}^{(t)} \right]^{-1}\\
&\times \left\{-\sum_{i=1}^n z_{ig}^{(t+1)}\left(x_{im}-\frac{1}{2}\right)\bs \mu_{ig}^{(t+1)}-\frac{\sum_{i=1}^{n} z_{ig}^{(t+1)}(x_{im}-\frac{1}{2})\sum_{i=1}^{n}z_{ig}^{(t+1)}B(\xi_{img}^{(t+1)})\bs \mu_{ig}^{(t+1)} }{\sum_{i=1}^{n}z_{ig}^{(t+1)}B(\xi_{img}^{(t+1)})-n_g^{(t+1)}} \right\},\\
&
\alpha_{mg}^{(t+1)}=-\left\{2\sum_{i=1}^n z_{ig}^{(t+1)} B(\xi_{img}^{(t+1)})-n_g^{(t+1)} \right\}^{-1}\\
&\;\;\;\;\;\;\;\;\;\;\;\;\;\;\left[\sum_{i=1}^n z_{ig}^{(t+1)} \left\{x_{im}-\frac{1}{2}+2B(\xi_{img}^{(t+1)})\big(\vecw_{mg}^{(t+1)}\big)'\bs \mu_{ig}^{(t+1)}\right\} \right],
%&\alpha_{mg}^{(t+1)}=-\left[2\sum_{i=1}^n z_{ig}^{(t+1)} B(\xi_{img}^{(t+1)})-n_g^{(t+1)} \right]^{-1}\left[\sum_{i=1}^n z_{ig}^{(t+1)} \left(x_{im}-\frac{1}{2}+2B(\xi_{img}^{(t+1)})\vecw_{mg}^{'(t+1)}\bs \mu_{ig}^{(t+1)}\right) \right],
\end{split}
\end{equation*}}where $n_g^{(t+1)}=\sum_{i=1}^nz_{ig}^{(t+1)}$, $\mathbb{E}(\vecY_{ig} \vecY'_{ig})=\bs \Sigma_{ig}^{(t+1)}+\bs \mu_{ig}^{(t+1)}\bs \mu_{ig}^{'(t+1)}$, $\bs \Gamma_{mg}^{(t)} =\text{diag}\left({1}/{|w_{1mg}^{(t)}|}, \ldots, {1}/{|w_{dmg}^{(t)}|}\right),$ and 

\begin{equation*}
\begin{split}
&\left\{\sum_{i=1}^n z_{ig}^{(t+1)} B(\xi_{img}^{(t+1)}) \mathbb{E}(\vecY_{ig})\right\}\left\{\sum_{i=1}^n z_{ig}^{(t+1)} B(\xi_{img}^{(t+1)}) \mathbb{E}(\vecY'_{ig})\right\}\\
&\ =\sum_{i=1}^n \big(z_{ig}^{(t+1)}\big)^2B\big\{\big(\xi_{img}^{(t+1)}\big)^2\big\}\mathbb{E}(\vecY_{ig} \vecY'_{ig})\\
&\ +2\sum_{i<j}\big(z_{ig}^{(t+1)}\big)^2\big(z_{jg}^{(t+1)}\big)^2B\big\{\big(\xi_{img}^{(t+1)}\big)^2\big\}B\big\{\big(\xi_{jmg}^{(t+1)}\big)^2\big\}\mathbb{E}(\vecY_{ig}\vecY'_{jg}).
\end{split}
\end{equation*}
We adopt a numerically more convenient form of the update for $\vecw_{mg}$:
{\small
\begin{equation*}
\begin{split}
&\vecw_{mg}^{(t+1)}=\bs \Upsilon_{mg}^{(t)}\left(\bs \Upsilon_{mg}^{(t)}\left[2\sum_{i=1}^n z_{ig}^{(t+1)}B(\xi_{img}^{(t+1)})\mathbb{E}(\vecY_{ig}\vecY'_{ig})\right.\right.\\
&\left.\left.+\frac{2\{\sum_{i=1}^n z_{ig}^{(t+1)} B(\xi_{img}^{(t+1)}) \mathbb{E}(\vecy_{ig})\}\{\sum_{i=1}^n z_{ig}^{(t+1)} B(\xi_{img}^{(t+1)}) \mathbb{E}(\vecY'_{ig})\}}{\sum_{i=1}^n z_{ig}^{(t+1)} B(\xi_{img}^{(t+1)})-n_g^{(t+1)}}\right]\bs \Upsilon_{mg}^{(t)}-\lambda_{mg}\mathbf{I}_{D}\right)^{-1}\\
&\times \bs \Upsilon_{mg}^{(t)}\left\{-\sum_{i=1}^n \left(x_{im}-\frac{1}{2}\right)\bs \mu_{ig}^{(t+1)}-\frac{\sum_{i=1}^{n} z_{ig}^{(t+1)}(x_{im}-\frac{1}{2})\sum_{i=1}^{n}z_{ig}^{(t+1)}B(\xi_{img}^{(t+1)})\bs \mu_{ig}^{(t+1)} }{\sum_{i=1}^{n}z_{ig}^{(t+1)}B(\xi_{img}^{(t+1)})-n_g^{(t+1)}} \right\},\\
\end{split}
\end{equation*}}
where $\bs \Upsilon_{mg}^{(t)}=\text{diag}\left(|w_{1mg}^{(t)}|^{1/2}, \ldots, |w_{dmg}^{(t)}|^{1/2}\right)$. This avoids estimating $|w_{dmg}^{-1(t)}|$, some of which are expected to go to zero. 
%%%%%%
%%%%%%
%%%%%%

\section{Tables} \label{sec:sim4k}
\begin{table}[h]
	\caption{\label{table:compare}Percent preferred by the BIC and average ARI with $G=1, \ldots, 6$.}	
	\centering
		\begin{tabular}{rccccccc}
		  \hline
 		&&\multicolumn{2}{c}{PMLTM} & \multicolumn{2}{c}{MLTA} & \multicolumn{2}{c}{LDA} \\ 
				\cline{3-4}\cline{5-6}\cline{7-8}
		               &&BIC&ARI*&BIC&ARI*&BIC&ARI\\
 		 \hline
\multirow{3}{*}{\hspace{0.2cm}$n=300$}&$G=2$&92&0.74&92&0.72&N/A&0.10\\
&$G=3$&92&0.72&85&0.70&N/A&0.15\\
&$G=5$&90&0.72&67&0.24&N/A&0.18\\
\hline
\multirow{3}{*}{$n=600$}&$G=2$&99&0.85&94&0.86&N/A&0.26\\
&$G=3$&97&0.83&89&0.84&N/A&0.20\\
&$G=5$&95&0.80&86&0.57&N/A&0.18\\
\hline
\multirow{3}{*}{$n=900$}&$G=2$&100&0.89&96&0.90&N/A&0.29\\
&$G=3$&100&0.85&94&0.85&N/A&0.30\\
&$G=5$&100&0.83&90&0.57&N/A&0.24\\
\hline
	\end{tabular}
\end{table}%

%%%%%%%

\begin{table}[h]
\caption{\label{ch6:table:4.5}Sample reviews of our chosen model ($G=4$, $d=2$).}
\centering
\begin{tabular*}{1\textwidth}{@{\extracolsep{\fill}}lp{6in}}
\hline 
Cltr.&Reviews \\ [0.5ex] % inserts table %heading
\hline \vspace{0.12in}
1 &1. ``The place is really well furnished, pleasant and clean. Islam was very helpful, you can feel free to ask him virtually
 anything and he'll help you. He was fun too, very cool talking to him. Oh, and the place is pretty conveniently located too. Highly recommended. The neighbourhood might not be the cleanest in Boston (my gf liked Brooklyne much more in that matter), but this is a great location and price for value overall.''\\
 &2. ``Perry's house is much cleaner and bigger than it is in the pictures. We are very happy to stay at his apartment. Perry is also very friendly and thoughtful. He explained all the instructions very clearly and he kept contacting us to know if we had any question. The house is located in a nice neighborhood, about 5 minute walking to a train/subway station.''\\ 
 &3. ``We stayed here for almost 2 months when we relocated to Boston quite quickly.  The apartment was very clean and very new. Perry went out of his way on multiple occasions to make sure that me, my husband and our 18 month old son had everything we needed.
The kitchen and bathroom are very newly renovated and the kitchen had everything we needed (appliances, pots/pans, etc).
We had a great experience here and would definitely recommend it.''\\ \hline
\end{tabular*}
\end{table}

\begin{table}
\caption{\label{ch6:table:4.5b}Sample reviews of our chosen model ($G=4$, $d=2$). (continued from Table~9)}
\centering
\begin{tabular*}{1\textwidth}{@{\extracolsep{\fill}}lp{6in}}
\hline 
Cltr.&Reviews \\ [0.5ex] % inserts table %heading
\hline \vspace{0.12in}
%1 &1. ``The place is really well furnished, pleasant and clean. Islam was very helpful, you can feel free to ask him virtually
% anything and he'll help you. He was fun too, very cool talking to him. Oh, and the place is pretty conveniently located too. Highly recommended. The neighbourhood might not be the cleanest in Boston (my gf liked Brooklyne much more in that matter), but this is a great location and price for value overall.''\\
% &2. ``Perry's house is much cleaner and bigger than it is in the pictures. We are very happy to stay at his apartment. Perry is also very friendly and thoughtful. He explained all the instructions very clearly and he kept contacting us to know if we had any question. The house is located in a nice neighborhood, about 5 minute walking to a train/subway station.''\\ 
% &3. ``We stayed here for almost 2 months when we relocated to Boston quite quickly.  The apartment was very clean and very new. Perry went out of his way on multiple occasions to make sure that me, my husband and our 18 month old son had everything we needed.
%The kitchen and bathroom are very newly renovated and the kitchen had everything we needed (appliances, pots/pans, etc).
%We had a great experience here and would definitely recommend it.''\\ \hline
2& 1. ``Izzy's communication is very good. All communication was done via text or AirBnB messaging. Directions and house details were well spelled out and clear. I was in the basement room of the 3 rooms he rents out. Everything is clean but spares. I would not consider it cozy but it was a very good value.''\\&2. ``We were rather disappointed with this accommodation. The host did not even meet us, but left rather complicated instructions to access the keys to the apartment.  We did not meet the host at all during our stay, or even hear from him as to how we were getting on.  The apartment was somewhat shabby, and not really like the image indicated, as this only showed a small corner of one room.  The kitchen was tiny, and although quite well equipped, it badly needed redecoration and a good clean. In addition, the apartment backed onto a yard with three dumpsters, and on 4 occasions we were awakened early in the morning by the noise of the dumpsters being emptied.''\\&3. ``I fell in love with the view of this apartment. Fenway out the window as promised. My expectations were pretty low going in because I realized it was very basic budget accommodations. Sean was helpful with the different questions I had about the city. The instructions for obtaining lockbox key were very clear. The location is great and the building old and had a lot of character. I came to town with a friend of mine for the night to catch the Red Sox game. We understood it to have a large enough bed to accommodate us since it says 1 to 4 people. When we arrived the bed seemed quite small. When I asked Sean about it he told me that there was 2 mattresses on top of each other and to take them apart and he thought that there were sheets in the closet for both  ( there were not) we had explored Boston all day and didn't return til 1 am..pulling a mattress apart was not what I wanted to do. We were so tired and since there was only 1 sheet we decided to just be very cozy. The bed was comfortable and we slept well until around 5 am when people were down in the alley going through glass bottles in the trash dumpsters which was very loud. (Not sure if that happens all the time) The kitchen is small but would be helpful if you needed one. I would not recommend having 4 people stay as it would be quite cramped ( but if you are looking for a budget place with a great view..this would work.)''\\  \hline
%3& 1. ``The host was pleasant and prompt in her responses. The apartment is spacious and tidy, however there is an unpleasant odor and the rugs, floor and furniture are full of dog hair. It ended up getting on our clothes. Also, the stairs leading up to the apt have dirt and dog hair and are cluttered.  The location is great if you are doing things in Jamaica Plain. There is an awesome place called The Frogmore a block away as well as a pizza place (that only takes cash) and Whole Foods is very close by.''\\&2. ``Everything about this studio was perfect! I couldn't have wished for a better first Airbnb experience! Thanks Katie!''\\&3. ``Lisa's place was good. When we got there the hallway was smelling bad. The apartment has a lot of space, but it wasn't really clean and we even find a pill on the floor. The kitchen could have more things, there was only 4 forks and knifes. Small plates and A LOT of frying pans and all didn't look clean. The beds were comfortable but the sheets and the apartment were smelly. We spent most of our days out of the apartment, if that's what you are going to do, this is a good place for you.''\\\hline
%4& 1. ``GREAT SPACE, PERFECT LOCATION, AWESOME PEOPLE!! Definately will be back!!!!''\\&2. ``We liked the apartment but not the three flights of steps to get to it.''\\&3. ``Everything was great - as described and expected.''\\[1ex]
%\hline
\end{tabular*}
\end{table}

\begin{table}
\caption{\label{ch6:table:4.5c}Sample reviews of our chosen model ($G=4$, $d=2$). (continued from Table~10)}
\centering
\begin{tabular*}{1\textwidth}{@{\extracolsep{\fill}}lp{6in}}
\hline 
Cltr.&Reviews \\ [0.5ex] % inserts table %heading
\hline \vspace{0.12in}
3& 1. ``The host was pleasant and prompt in her responses. The apartment is spacious and tidy, however there is an unpleasant odor and the rugs, floor and furniture are full of dog hair. It ended up getting on our clothes. Also, the stairs leading up to the apt have dirt and dog hair and are cluttered.  The location is great if you are doing things in Jamaica Plain. There is an awesome place called The Frogmore a block away as well as a pizza place (that only takes cash) and Whole Foods is very close by.''\\&2. ``Everything about this studio was perfect! I couldn't have wished for a better first Airbnb experience! Thanks Katie!''\\&3. ``Lisa's place was good. When we got there the hallway was smelling bad. The apartment has a lot of space, but it wasn't really clean and we even find a pill on the floor. The kitchen could have more things, there was only 4 forks and knifes. Small plates and A LOT of frying pans and all didn't look clean. The beds were comfortable but the sheets and the apartment were smelly. We spent most of our days out of the apartment, if that's what you are going to do, this is a good place for you.''\\\hline
4& 1. ``GREAT SPACE, PERFECT LOCATION, AWESOME PEOPLE!! Definately will be back!!!!''\\&2. ``We liked the apartment but not the three flights of steps to get to it.''\\&3. ``Everything was great - as described and expected.''\\[1ex]
\hline
\end{tabular*}
\end{table}

\clearpage
\section{Results from Experiment 1}\label{app:addtables}
The results from Experiment 1 (Section~3.5.2) are summarized in Tables~\ref{table:time} and~\ref{sim1result}.
Table~\ref{table:time} shows a comparison of the average run time over 100 loops of the VE-step and M-steps  using {\sf R} and {Python} ($G=2$, $D=3$, $n=900$). {Python} runs approximately 103 times faster than {\sf R} for the VE-step and 190 times faster for the M-steps. Table~\ref{sim1result} shows the average model parameter estimates as well as their standard deviations for the first simulation experiment.
\begin{table}[h]
\caption{\label{table:time}A comparison of run times for {\sf R} and Python based on simulated data.}
\centering
\begin{tabular*}{1\textwidth}{@{\extracolsep{\fill}} rccc}
\hline
\hspace{0.2cm}Function&Number of Loops& Python&{\sf R}\\
\hline
VE-Step&100& 15.4 ms/loop&1.7 s/loop\\
M-Steps&100& 4.19 ms/loop&0.9 s/loop\\ [1ex]
\hline
\end{tabular*}
\end{table}

\begin{landscape}
\small{
\begin{longtable}{lll}
\caption{Average model parameter estimates as well as their standard deviations from the 100 runs for the first simulation experiment.}
\label{sim1result}\\
                \hline
		\multicolumn{3}{c}{$n=300$}\\
		\hline
	\multirow{2}{*}{$\bs\alpha_1$}&Mean&$(-2.1,-2.9,-1.6,-2.5,-5.9,-3.8,-4.3,-2.4,-0.1,-0.4,-4.5,2.3,-5.3,-3.9,-2.0,0.9,0.5,-1.5,-2.0,-2.2)$\\
        &Std.D.&$(0.8,0.1,0.5,0.2,0.5,0.5,0.7,0.5,0.5,0.4,0.0,0.1,0.8,0.4,0.3,0.7,0.8,0.4,0.7,0.9)$\\
	\multirow{2}{*}{$\bs\alpha_2$}&Mean&$(4.5,2.8,0.5,2.0,1.6,0.4,3.4,1.2, 5.9, 3.9,1.7,3.4,-0.9,0.6,-1.3,3.6,1.5,3.4,0.9,6.7)$\\
        &Std.D.&$(0.9,0.8,0.1,0.0,0.2,0.1,0.3,0.4,0.5,0.5,0.7,0.5,0.1,0.8,0.3,0.5,0.5,0.3,0.2,1.0)$\\	
	\multirow{2}{*}{$\vecw_1$}&Mean&$\left[\arraycolsep=1.5pt\def\arraystretch{1}\begin{array}{cccccccccccccccccccc}-1.6&0&-2.5&-0.8&-0.6&-1.7&-0.6&0&-0.9&-1.1&-2.9&0.2&-2.5&-0.3&-2.3&-3.2&-0.7&-3.1&-0.1&-0.7\\
	-0.1&-0.7&-0.1&-0.7&0&-0.1&-0.6&-0.7&-0.8&-0.5&-0.4&-0.1&0&0&-0.6&-0.3&-0.2&-0.7&-0.1&-0.6\\
	-1.2&-0.2&-1.1&0.5&0&-1.3&0.5&0.5&0.4&0.5&0.5&-0.1&-0.6&-0.9&-0.1&0&-0.1&-0.8&-0.2&0\end{array}\right]$\\
	&Std.D.&$\left[\arraycolsep=1.5pt\def\arraystretch{1}\begin{array}{cccccccccccccccccccc}
	0.5&0.5&0.2&0.3&0.4&0.8&1.0&0.6&0.5&0.4&0.7&0.7&0.6&0.5&0.6&0.5&0.6&0.8&0.4&0.4\\
	0.4&0.8&0.2&0.4&0.7&0.4&0.5&0.5&0.3&0.6&0.4&0.4&0.3&0.4&0.9&0.6&0.3&0.3&0.4&0.3\\
	0.6&0.1&0.3&0.6&0.5&0.3&0.2&0.4&0.1&0.2&0.5&0.5&0.3&0.2&0.4&0.6&0.5&0.3&0.5&0\end{array}\right]$\\
		\multirow{2}{*}{$\vecw_2$}&Mean&$\left[\arraycolsep=1.5pt\def\arraystretch{1}\begin{array}{cccccccccccccccccccc}		
		1.0&0.5&1.6&-0.9&-0.7&1.3&0&-0.1&-0.5&0&-0.7&-0.3&0.6&1.3&1.1&0&-0.2&0.4&0&-0.2\\
		0.4&0.9&0.8&1.0&0&0.5&0.1&0.8&0&1.3&1.0&0.9&0&0.8&0.8&0.1&1.0&0&0.7&0\\
		0&0.9&0.2&0&-0.8&0&0&0.3&-0.4&-0.3&-0.9&-1.5&0.5&0&0.9&-1.0&-1.1&0&-1.2&-0.6\end{array}\right]$\\
	&Std.D.&$\left[\arraycolsep=1.5pt\def\arraystretch{1}\begin{array}{cccccccccccccccccccc}
	0.2&0.2&0.3&0.4&0.5&0.2&0.2&0.7&0.3&0.5&0.8&0.4&0.3&0.6&0.7&0.5&0.6&0.3&0.2&0.2\\
	0.3&0.3&0.4&0.6&0.4&0.2&0.1&0.1&0.5&0.2&0.4&0.5&0.3&0.2&0.5&0.3&0.2&0.6&0.1&0.3\\
        0.3&0.2&0.3&0.5&0.1&0.3&0.2&0.2&0.2&0.3&0.6&0.6&0.6&0.3&0.5&0.4&0.4&0.3&0.4&0.2\end{array}\right]$\\
	$\bs\pi$&Mean [Std.D.]&(0.54,0.46) [0.03,0.03]\\
	ARI&Mean [Std.D.] &0.72 [0.09] \\
	\hline
	\multicolumn{3}{c}{$n=600$}\\
	\hline
	\multirow{2}{*}{$\bs\alpha_1$}&Mean&$(-1.3,-2.2,-1.3,-2.0,-4.9,-3.6,-4.3,-2.6,-0.1,-0.6,-4.5,3.5,-4.9,-4.7,-2.7,1.2,0.3,-0.9,-1.5,-1.7)$\\
	&Std.D.&$(0.7,0.1,0.5,0.4,0.4,0.5,0.3,0.3,0.2,0.4,0.0,0.1,0.6,0.1,0.3,0.3,0.8,0.4,0.7,0.4)$\\
	\multirow{2}{*}{$\bs\alpha_2$}&Mean&$(5.6,3.2,0.2,2.2,2.3,0.2,3.5,1.5,6.0,4.5,2.1,3.7,-0.8,0.5,-1.5,4.1,1.3,3.6,0.9,7.4)$\\
	&Std.D.&$(0.7,0.7,0.1,0.0,0.2,0.1,0.3,0.4,0.5,0.1,0.7,0.6,0.1,0.7,0.5,0.4,0.5,0.3,0.2,0.6)$\\
	\multirow{2}{*}{$\vecw_1$}&Mean&$\left[\arraycolsep=1.5pt\def\arraystretch{1}\begin{array}{cccccccccccccccccccc}
	-0.8&0&-2.5&-0.8&-0.4&-0.8&-0.6&0&-1.0&-1.2&-2.9&0.4&-2.7&-0.2&-2.3&-3.4&-0.5&-2.9&0&-0.6\\
	-0.2&-0.8&0&-0.6&0&0&-0.8&-0.5&-0.7&-0.3&-0.3&0&0&0&-0.4&-0.2&-0.5&-0.4&-0.1&-0.6\\
	-1.1&-0.3&-1.5&0.5&0&-1.3&0.5&0.5&0.4&0.5&0.5&-0.1&-0.6&-0.9&-0.1&0&-0.1&-0.8&-0.2&0\end{array}\right]$\\
	&Std.D.&$\left[\arraycolsep=1.5pt\def\arraystretch{1}\begin{array}{cccccccccccccccccccc}
	0.2&0.5&0.2&0.3&0.4&0.8&0.8&0.6&0.3&0.4&0.7&0.7&0.6&0.5&0.6&0.5&0.6&0.4&0.4&0.1\\
	0.4&0.6&0.2&0.4&0.7&0.4&0.4&0.5&0.3&0.6&0.4&0.4&0.3&0.4&0.9&0.3&0.3&0.3&0.4&0.2\\
	0.3&0.1&0.3&0.5&0.3&0.3&0.2&0.3&0.1&0.2&0.2&0.5&0.3&0.2&0.4&0.4&0.5&0.3&0.3&0\end{array}\right]$\\
	\multirow{2}{*}{$\vecw_2$}&Mean&$\left[\arraycolsep=1.5pt\def\arraystretch{1}\begin{array}{cccccccccccccccccccc}		
		1.0&0.3&1.4&-0.8&-0.5&1.0&0&-0.1&-0.5&0&-0.3&-0.4&0.4&1.3&1.3&0&-0.3&0.5&0&-0.3\\
		0.6&0.9&0.9&1.1&0&0.7&0.1&0.7&0&1.0&1.2&0.4&0&0.6&0.4&0.1&1.0&0&0.7&0\\
		0&0.6&0.5&0&-0.9&0&0&0&-0.6&-0.4&-0.6&-1.5&0.6&0&0.9&-0.7&-1.2&0&-1.2&-0.5\end{array}\right]$\\
	&Std.D.&$\left[\arraycolsep=1.5pt\def\arraystretch{1}\begin{array}{cccccccccccccccccccc}
	0.2&0.2&0.3&0.4&0.1&0.2&0.2&0.7&0.3&0.4&0.8&0.4&0.3&0.6&0.7&0.4&0.6&0.3&0.2&0.2\\
	0.3&0.3&0.4&0.6&0.5&0.2&0.1&0.1&0.5&0.2&0.2&0.5&0.3&0.2&0.5&0.3&0.2&0.2&0.1&0.3\\
        0.3&0.2&0.4&0.2&0.1&0.3&0.2&0.2&0.2&0.3&0.3&0.6&0.5&0.3&0.3&0.4&0.3&0.3&0.4&0.1\end{array}\right]$\\
	$\bs\pi$&Mean [Std.D.]&(0.53,0.47) [0.03,0.03]\\
        ARI&Mean [Std.D.] &0.80 [0.09] \\
\hline
	\multicolumn{3}{c}{$n=900$}\\
	\hline
	\multirow{2}{*}{$\bs\alpha_1$}&Mean&$(-1.3,-2.2,-1.3,-2.0,-4.9,-4.6,-4.3,-2.0,-0.1,-0.7,-4.8,4.0,-5.3,-4.3,-2.0,1.6,0.3,-1.2,-1.5,-1.7)$\\
	&Std.D.&$(0.2,0.1,0.2,0.2,0.4,0.2,0.2,0.2,0.1,0.3,0.0,0.1,0.7,0.1,0.2,0.2,0.3,0.3,0.1,0.3)$\\
	\multirow{2}{*}{$\bs\alpha_2$}&Mean&$(6.0,3.4,0.1,2.2,1.8,0.4,4.0,1.6,5.5,4.9,1.9,3.7,-0.9,0.4,-1.6,3.8,1.1,3.9,0.7,6.3)$\\	
	&Std.D.&$(0.3,0.4,0.1,0.0,0.2,0.1,0.1,0.1,0.4,0.3,0.2,0.3,0.2,0.1,0.2,0.1,0.2,0.3,0.2,0.6)$\\
	\multirow{2}{*}{$\vecw_1$}&Mean&$\left[\arraycolsep=1.5pt\def\arraystretch{1}\begin{array}{cccccccccccccccccccc}
	-0.6&0&-2.0&-0.7&-0.4&-0.6&-0.7&0&-1.3&-1.1&-2.5&0.7&-2.1&-0.2&-2.4&-3.6&-0.7&-2.7&0&-0.5\\
	-0.2&-1.0&0&-0.6&0&0&-0.5&-0.5&-0.4&-0.1&-0.3&0&0&0&-0.7&-0.5&-0.2&-0.4&-0.1&-0.8\\
	-1.0&-0.5&-1.0&0.6&0&-1.5&0.6&0.5&0.3&0.8&0.3&0&-0.7&-0.6&-0.1&0&-0.1&-0.7&-0.2&0\end{array}\right]$\\
	&Std.D.&$\left[\arraycolsep=1.5pt\def\arraystretch{1}\begin{array}{cccccccccccccccccccc}
	0.1&0.5&0.2&0.2&0.2&0.7&0.8&0.6&0.2&0.3&0.7&0.7&0.4&0.5&0.6&0.3&0.6&0.4&0.3&0\\
	0.4&0.5&0.2&0.3&0.5&0.2&0.3&0.4&0.3&0.6&0.4&0.4&0.1&0.3&0.9&0.1&0.2&0.3&0.4&0\\
	0.1&0.0&0.3&0.5&0.2&0.2&0.2&0.1&0.1&0.1&0.2&0.3&0.1&0.2&0.2&0.4&0.5&0.2&0.2&0\end{array}\right]$\\
	\multirow{2}{*}{$\vecw_2$}&Mean&$\left[\arraycolsep=1.5pt\def\arraystretch{1}\begin{array}{cccccccccccccccccccc}		
		1.0&0.3&1.5&-0.8&-0.5&1.0&0&-0.1&-0.7&0&-0.2&-0.3&0.4&1.6&1.5&0&-0.2&0.5&0&-0.2\\
		0.7&0.8&1.0&1.4&0&0.9&0.0&0.6&0&1.3&1.2&0.4&0&0.4&0.4&0.1&1.0&0&0.6&0\\
		0&0.5&0.6&0&-0.7&0&0&0&-0.8&-0.4&-0.4&-1.9&0.6&0&1.0&-1.0&-1.1&0&-1.5&-0.4\end{array}\right]$\\
	&Std.D.&$\left[\arraycolsep=1.5pt\def\arraystretch{1}\begin{array}{cccccccccccccccccccc}
	0.1&0.2&0.1&0.3&0.0&0.2&0.1&0.7&0.2&0.4&0.8&0.2&0.2&0.6&0.5&0.4&0.5&0.3&0.1&0.1\\
	0.3&0.2&0.4&0.6&0.2&0.3&0.1&0.1&0.4&0.2&0.2&0.4&0.2&0.2&0.4&0.3&0.1&0.2&0.1&0.3\\
        0.2&0.2&0.3&0.2&0.1&0.2&0.1&0.2&0.2&0.2&0.3&0.4&0.3&0.3&0.3&0.2&0.3&0.1&0.4&0.1\end{array}\right]$\\
	$\bs\pi$&Mean [Std.D.]&(0.50,0.50) [0.01,0.01]\\
        ARI&Mean [Std.D.] &0.89 [0.05] \\
	\hline
\end{longtable}}
\end{landscape}}


\begin{thebibliography}{31}
\expandafter\ifx\csname natexlab\endcsname\relax\def\natexlab#1{#1}\fi
\expandafter\ifx\csname url\endcsname\relax
  \def\url#1{\texttt{#1}}\fi
\expandafter\ifx\csname urlprefix\endcsname\relax\def\urlprefix{URL: }\fi

\bibitem[{Aitken(1926)}]{aitken26}
Aitken, A.~C. (1926) A series formula for the roots of algebraic and
  transcendental equations.
\newblock \textit{Proceedings of the Royal Society of Edinburgh}, \textbf{45},
  14--22.

\bibitem[{Bird and Loper(2004)}]{bird04}
Bird, S. and Loper, E. (2004) {NLTK}: the natural language toolkit.
\newblock In \textit{Proceedings of the ACL 2004 on Interactive poster and
  demonstration sessions}, 31. Association for Computational Linguistics.

\bibitem[{Blei et~al.(2003)Blei, Ng and Jordan}]{blei03}
Blei, D.~M., Ng, A.~Y. and Jordan, M.~I. (2003) Latent dirichlet allocation.
\newblock \textit{Journal of Machine Learning Research}, \textbf{3}, 993--1022.

\bibitem[{B{\"o}hning et~al.(1994)B{\"o}hning, Dietz, Schaub, Schlattmann and
  Lindsay}]{bohning94}
B{\"o}hning, D., Dietz, E., Schaub, R., Schlattmann, P. and Lindsay, B.~G.
  (1994) The distribution of the likelihood ratio for mixtures of densities
  from the one-parameter exponential family.
\newblock \textit{Annals of the Institute of Statistical Mathematics},
  \textbf{46}, 373--388.

\bibitem[{Browne and McNicholas(2012)}]{browne12}
Browne, R.~P. and McNicholas, P.~D. (2012) Model-based clustering,
  classification, and discriminant analysis of data with mixed type.
\newblock \textit{Journal of Statistical Planning and Inference}, \textbf{142},
  2976--2984.

\bibitem[{Chang and Blei(2009)}]{chang09}
Chang, J. and Blei, D. (2009) Relational topic models for document networks.
\newblock In \textit{Artificial Intelligence and Statistics}, 81--88.

\bibitem[{Cheng and Jin(2019)}]{cheng19}
Cheng, M. and Jin, X. (2019) What do {A}irbnb users care about? an analysis of
  online review comments.
\newblock \textit{International Journal of Hospitality Management},
  \textbf{76}, 58--70.

\bibitem[{DeSantis et~al.(2008)DeSantis, Houseman, Coull, Stemmer-Rachamimov
  and Betensky}]{desantis08}
DeSantis, S.~M., Houseman, E.~A., Coull, B.~A., Stemmer-Rachamimov, A. and
  Betensky, R.~A. (2008) A penalized latent class model for ordinal data.
\newblock \textit{Biostatistics}, \textbf{9}, 249--262.

\bibitem[{Fan and Li(2001)}]{fan01}
Fan, J. and Li, R. (2001) Variable selection via nonconcave penalized
  likelihood and its oracle properties.
\newblock \textit{Journal of the American statistical Association},
  \textbf{96}, 1348--1360.

\bibitem[{Feinerer and Hornik(2015)}]{feinerer15}
Feinerer, I. and Hornik, K. (2015) \textit{tm: A Framework for Text Mining
  Applications within R}.
\newblock R package version 0.7-1.

\bibitem[{Gollini and Murphy(2014)}]{gollini14}
Gollini, I. and Murphy, T.~B. (2014) Mixture of latent trait analyzers for
  model-based clustering of categorical data.
\newblock \textit{Statistics and Computing}, \textbf{24}, 569--588.

\bibitem[{Guttentag et~al.(2018)Guttentag, Smith, Potwarka and
  Havitz}]{guttentag18}
Guttentag, D., Smith, S., Potwarka, L. and Havitz, M. (2018) Why tourists
  choose {A}irbnb: {A} motivation-based segmentation study.
\newblock \textit{Journal of Travel Research}, \textbf{57}, 342--359.

\bibitem[{He et~al.(2013)He, Zha and Li}]{he13}
He, W., Zha, S. and Li, L. (2013) Social media competitive analysis and text
  mining: A case study in the pizza industry.
\newblock \textit{International Journal of Information Management},
  \textbf{33}, 464--472.

\bibitem[{Houseman et~al.(2007)Houseman, Marsit, Karagas and Ryan}]{houseman07}
Houseman, E.~A., Marsit, C., Karagas, M. and Ryan, L.~M. (2007) Penalized item
  response theory models: application to epigenetic alterations in bladder
  cancer.
\newblock \textit{Biometrics}, \textbf{63}, 1269--1277.

\bibitem[{Hubert and Arabie(1985)}]{hubert85}
Hubert, L. and Arabie, P. (1985) Comparing partitions.
\newblock \textit{Journal of Classification}, \textbf{2}, 193--218.

\bibitem[{Jaakkola and Jordan(2000)}]{jaakkola00}
Jaakkola, T.~S. and Jordan, M.~I. (2000) Bayesian parameter estimation via
  variational methods.
\newblock \textit{Statistics and Computing}, \textbf{10}, 25--37.

\bibitem[{Knott and Bartholomew(1999)}]{knott99}
Knott, M. and Bartholomew, D.~J. (1999) \textit{Latent variable models and
  factor analysis}.
\newblock No.~7. London: Edward Arnold.

\bibitem[{Lafferty and Blei(2006)}]{lafferty06}
Lafferty, J.~D. and Blei, D.~M. (2006) Correlated topic models.
\newblock In \textit{Advances in neural information processing systems},
  147--154.

\bibitem[{Mazumder et~al.(2012)Mazumder, Friedman and Hastie}]{mazumder12}
Mazumder, R., Friedman, J.~H. and Hastie, T. (2012) Sparsenet: Coordinate
  descent with nonconvex penalties.
\newblock \textit{Journal of the American Statistical Association},
  \textbf{106}, 1125--1138.

\bibitem[{McLachlan and Peel(2000)}]{mclachlan00}
McLachlan, G. and Peel, D. (2000) \textit{Finite Mixture Models}.
\newblock New York: John Wiley \& Sons.

\bibitem[{Muthen et~al.(2006)Muthen, Asparouhov et~al.}]{muthen06}
Muthen, B., Asparouhov, T. et~al. (2006) Item response mixture modeling:
  Application to tobacco dependence criteria.
\newblock \textit{Addictive Behaviors}, \textbf{31}, 1050--1066.

\bibitem[{Park and Casella(2008)}]{park08}
Park, T. and Casella, G. (2008) The {B}ayesian {L}asso.
\newblock \textit{Journal of the American Statistical Association},
  \textbf{103}, 681--686.

\bibitem[{Rand(1971)}]{rand71}
Rand, W.~M. (1971) Objective criteria for the evaluation of clustering methods.
\newblock \textit{Journal of the American Statistical Association},
  \textbf{66}, 846--850.

\bibitem[{Schwarz(1978)}]{schwarz78}
Schwarz, G. (1978) Estimating the dimension of a model.
\newblock \textit{The Annals of Statistics.}, \textbf{6}, 461--464.

\bibitem[{Small and Harris(2014)}]{small14}
Small, J. and Harris, C. (2014) Crying babies on planes: Aeromobility and
  parenting.
\newblock \textit{Annals of Tourism Research}, \textbf{48}, 27--41.

\bibitem[{Taddy(2013)}]{taddy13}
Taddy, M. (2013) Multinomial inverse regression for text analysis.
\newblock \textit{Journal of the American Statistical Association},
  \textbf{108}, 755--770.

\bibitem[{Tibshirani(1996)}]{tibshirani96}
Tibshirani, R. (1996) Regression shrinkage and selection via the {L}asso.
\newblock \textit{Journal of the Royal Statistical Society. Series B
  (Statistical Methodology)}, \textbf{58}, 267--288.

\bibitem[{Tipping(1999)}]{tipping992}
Tipping, M.~E. (1999) Probabilistic visualisation of high-dimensional binary
  data.
\newblock In \textit{Advances in Neural Information Processing Systems 11}
  (eds. M.~J. Kearns, S.~A. Solla and D.~A. Cohn), 592--598. Cambridge: MIT
  Press.

\bibitem[{Tussyadiah and Zach(2017)}]{tussyadiah17}
Tussyadiah, I.~P. and Zach, F. (2017) Identifying salient attributes of
  peer-to-peer accommodation experience.
\newblock \textit{Journal of Travel \& Tourism Marketing}, \textbf{34},
  636--652.

\bibitem[{Vermunt(2007)}]{vermunt07}
Vermunt, J.~K. (2007) Multilevel mixture item response theory models: An
  application in education testing.
\newblock In \textit{Proceedings of the 56th session of the International
  Statistical Institute.}, 22--28. Lisbon, Portugal.

\bibitem[{Yuan and Wei(2014)}]{yuan14}
Yuan, J. and Wei, G. (2014) An efficient {M}onte {C}arlo {EM} algorithm for
  {B}ayesian {L}asso.
\newblock \textit{Journal of Statistical Computation and Simulation},
  \textbf{84}, 2166--2186.

\end{thebibliography}
\end{document}